\documentclass[twocolumn,showpacs,aps,floatfix,prd]{revtex4}

\usepackage{graphicx}
\usepackage{dcolumn}
\usepackage{amsmath}
\usepackage{epsfig}
\usepackage{hhline}

\RequirePackage{xspace}

\newcommand{\BABARPubYear}    {02}
\newcommand{\BABARPubNumber}  {012}
\newcommand{\SLACPubNumber} {9579}
\usepackage{relsize}
\def\babar{\mbox{\slshape B\kern-0.1em{\smaller A}\kern-0.1em
    B\kern-0.1em{\smaller A\kern-0.2em R}}\xspace}
\def\pep2{PEP-II}
\def\Kz {\ensuremath{K^0}\xspace}
\def\Kbar    {\kern 0.18em\overline{\kern -0.18em K}{}\xspace}
\def\Kzbar {\ensuremath{\Kbar^0}\xspace}
\def\Dbar    {\kern 0.18em\overline{\kern -0.18em D}{}\xspace}
\def\Dzbar {\ensuremath{\Dbar^0}\xspace}
\def\Bz {\ensuremath{B^0}\xspace}
\def\Bbar    {\kern 0.18em\overline{\kern -0.18em B}{}\xspace}
\def\Bzbar {\ensuremath{\Bbar^0}\xspace}
\def\BdashBbar {\Bz-\Bzbar}
\def\BzBzbar {\ensuremath{\Bz\Bzbar}\xspace}
\def\tauBz      {\ensuremath{\tau_{B^0}}\xspace}
\def\tauBp      {\ensuremath{\tau_{B^+}}\xspace}
\def\dm  {\ensuremath{\delta m}\xspace}
\def\Dm		{\ensuremath{\Delta m_{d}}\xspace}
\def\Dt		{\ensuremath{\Delta t}\xspace}
\def\sigmaDt {\ensuremath{\sigma_{\Dt}}\xspace}
\def\sigmaz {\ensuremath{\sigma_{z}}\xspace}
\def\Dttrue		{\ensuremath{{\Delta t}_{\rm true}}\xspace}

\def\dDt {\ensuremath{\delta\Delta t}\xspace}
\def\Dz		{\ensuremath{\Delta z}\xspace}
\def\ks  {\ensuremath{K^0_{\scriptscriptstyle S}}\xspace}
\def\FourS {\ensuremath{\Upsilon(4S)}\xspace}
\def\dz {\ensuremath{D^0}\xspace}
\def\dst {\ensuremath{D^{*}}\xspace}

\def\dstm {\ensuremath{D^{*-}}\xspace}

\def\btodstlnu  {\ensuremath{B^0 \rightarrow D^{*-}\ell^+\nu_\ell}\xspace}
\def\dstlnu  {\ensuremath{D^{*-}\ell^+\nu_\ell}\xspace}
\def\dstl  {\ensuremath{D^{*-}\ell^+}\xspace}
\def\dste  {\ensuremath{D^{*-}e^+}\xspace}
\def\dstmu  {\ensuremath{D^{*-}\mu^+}\xspace}
\def\dsttodpi  {\ensuremath{D^{*-}\rightarrow \Dzbar \pi^-}\xspace}
\def\kpi  {\ensuremath{K^+\pi^-}\xspace}
\def\kpipiz  {\ensuremath{K^+\pi^-\pi^0}\xspace}
\def\kpipipi  {\ensuremath{K^+\pi^-\pi^+\pi^-}\xspace}
\def\kspipi  {\ensuremath{\ks\pi^+\pi^-}\xspace}
\def\tdstl {\ensuremath{t_{D^*\ell}}\xspace}
\def\ttag  {\ensuremath{t_{\rm tag}}\xspace}

\def\massdiff  {\ensuremath{m(D^*) - m(D^0)}\xspace}
\def\thby {\ensuremath{\theta_{B,\dst\ell}}\xspace}
\def\thbyfl {\ensuremath{\theta_{B,\dst(-\ell)}}\xspace}
\def\costhby {\ensuremath{\cos\thby}\xspace}
\def\costhbyfl {\ensuremath{\cos\thbyfl}\xspace}
\def\zdstl {\ensuremath{z_{D^*\ell}}\xspace}
\def\ztag {\ensuremath{z_{\rm tag}}\xspace}
\def\bdstl {\ensuremath{B_{D^*\ell}}\xspace}
\def\btag {\ensuremath{B_{\rm tag}}\xspace}
\def\chisq {\ensuremath{\chi^2}\xspace}
\def\lepton {\ensuremath{{\tt lepton}}\xspace}
\def\kaon {\ensuremath{{\tt kaon}}\xspace}
\def\ntone {\ensuremath{{\tt NT1}}\xspace}
\def\nttwo {\ensuremath{{\tt NT2}}\xspace}
\def\ntthree {\ensuremath{{\tt NT3}}\xspace}
\def\wBp {\ensuremath{\omega_{B^+}}\xspace}
\def\wBz {\ensuremath{\omega_{B^0}}\xspace}
\def\fBp {\ensuremath{f_{B^+}}\xspace}

\newcommand{\GeV}{\ensuremath{\mathrm{Ge\kern -0.1em V}}\xspace}
\newcommand{\MeV}{\ensuremath{\mathrm{Me\kern -0.1em V}}\xspace}
\newcommand{\keV}{\ensuremath{\mathrm{ke\kern -0.1em V}}\xspace}
\newcommand{\eV}{\ensuremath{\mathrm{e\kern -0.1em V}}\xspace}
\newcommand{\GeVc}{\ensuremath{\mathrm{Ge\kern -0.1em V}\kern -0.1em /c}\xspace}
\newcommand{\MeVc}{\ensuremath{\mathrm{Me\kern -0.1em V}\kern -0.1em /c}\xspace}
\newcommand{\GeVcc}{\ensuremath{\mathrm{Ge\kern -0.1em V}\kern -0.1em /c^2}\xspace}
\newcommand{\MeVcc}{\ensuremath{\mathrm{Me\kern -0.1em V}\kern -0.1em /c^2}\xspace}

\long\def\inst#1{\par\nobreak\kern 4pt\nobreak
  {\it #1}\par\vskip 10pt plus 3pt minus 3pt}

\begin{document}

\begin{flushleft}
\babar-PUB-\BABARPubYear/\BABARPubNumber \\
SLAC-PUB-\SLACPubNumber \\
\end{flushleft}

% Title of the paper
\title{\large \bf
\Large
Simultaneous Measurement of the $B^0$ Meson Lifetime and Mixing Frequency 
with $B^0 \rightarrow D^{*-}\ell^+\nu_\ell$ Decays
\begin{center}
\vskip 5mm
The \babar Collaboration
\end{center}
} % end title{}

% Input author list file
%% author list as of 01-Oct-2002 (559 authors)
%
\author{B.~Aubert}
\author{R.~Barate}
\author{D.~Boutigny}
\author{J.-M.~Gaillard}
\author{A.~Hicheur}
\author{Y.~Karyotakis}
\author{J.~P.~Lees}
\author{P.~Robbe}
\author{V.~Tisserand}
\author{A.~Zghiche}
\affiliation{Laboratoire de Physique des Particules, F-74941 Annecy-le-Vieux, France }
% ...
\collaboration{The \babar\ Collaboration}
\noaffiliation

%\today
\date{\today}

% Abstract
\begin{abstract}
We measure the $B^0$ lifetime $\tau_{B^0}$ and 
the $B^0$-${\kern 0.18em\overline{\kern -0.18em B}}{}^0$
oscillation frequency $\Delta m_d$
with a sample of approximately 14,000 exclusively reconstructed
$B^0 \rightarrow D^{*-}\ell^+\nu_\ell$ signal events, 
selected from 23 million 
$B\kern 0.18em\overline{\kern -0.18em B}$ pairs recorded
at the $\Upsilon(4S)$ resonance with the 
${\mbox{\slshape B\kern-0.1em{\smaller A}\kern-0.1em
 B\kern-0.1em{\smaller A\kern-0.2em R}}\xspace}$
detector at the Stanford Linear Accelerator Center. 
The decay position of the other $B$ is determined with the remaining
tracks in the event, and its $b$-quark flavor at the time of decay is
determined with a tagging algorithm that exploits the correlation
between the flavor of the $b$-quark and the charges of its decay
products.
The lifetime and oscillation frequency are measured 
simultaneously with an unbinned maximum-likelihood fit that uses, 
for each event, the measured difference in decay times of the two $B$
mesons ($\Delta t$), the calculated uncertainty on $\Delta t$,
the signal and background probabilities,
and $b$-quark tagging information for the other $B$.
The results are 
$$
\tau_{B^0} = (1.523^{+0.024}_{-0.023} \pm 0.022)~\rm{ps}
$$
and
$$
\Delta m_d = (0.492 \pm 0.018 \pm 0.013)~\rm{ps}^{-1}.
$$
The statistical correlation coefficient between $\tau_{B^0}$ and 
$\Delta m_d$ is $-0.22$.

\end{abstract}

% PACS, the Physics and Astronomy Classification Scheme.

\maketitle

\newpage

%%%%%%%%%%%%%%%%%%%%%%%%%%%%%%%%%%%%%%%%%%%%%

% The body of the paper starts here
\setcounter{footnote}{0}

%%%%%%%%%%%%%%%%%%%%%%%%%%%%%%%%%%%%%%%%%%%%%

\section{Introduction and analysis overview}

\label{sec:intro}

The time evolution of $B^0$ mesons is governed by the overall
decay rate $1/\tauBz$ and the \BdashBbar oscillation frequency
\Dm.
The phenomenon of particle-antiparticle oscillations or ``mixing'' 
has been observed in neutral mesons containing a down quark 
and either a strange quark ($K$ mesons)~\cite{ref:Kmix}
or a bottom quark ($B$ mesons)~\cite{ref:Bmix}.
In the Standard Model of particle physics, mixing is the result 
of second-order charged weak interactions involving box diagrams 
containing virtual quarks with charge $2/3$.
In $B$ mixing, the diagrams containing the top quark dominate due to
the large mass of the top quark.
Therefore, the mixing frequency is sensitive to the 
Cabibbo-Kobayashi-Maskawa quark-mixing matrix element 
$V_{td}$~\cite{ref:CKM}.
In the neutral $K$ meson system, mixing also has contributions from
real intermediate states accessible to both a $\Kz$ and 
a $\Kzbar$ meson. Real intermediate states lead to a difference in
the decay rate for the two mass eigenstates of the neutral meson
system. 
For the $B$ system, the decay rate difference is expected to be of 
${\cal O}(10^{-2}\text{--}10^{-3})$ times smaller~\cite{ref:bigi} than the
average decay rate and the mixing frequency, and is ignored in this
analysis. 

We present a simultaneous measurement of the \Bz lifetime and
oscillation frequency based on a sample of approximately 14,000 
exclusively reconstructed \btodstlnu decays~\cite{cc}
selected from a 
sample of 23 million $B\overline B$ events recorded at the \FourS resonance
with the \babar detector~\cite{ref:babar} at the Stanford Linear
Accelerator Center, in 1999--2000.
In this experiment, 9-\GeV electrons and 3.1-\GeV positrons, 
circulating in the PEP-II storage ring~\cite{ref:pepii},
annihilate
to produce $B\overline B$ pairs moving along the $e^-$ beam
direction ($z$ axis) with a Lorentz boost of $\beta\gamma = 0.55$,
allowing a measurement of the proper time difference between the
two $B$ decays, $\Dt$.

The decay-time difference $\Dt$ between two neutral $B$
mesons produced in a coherent $P$-wave state in an \FourS event
is governed by the probabilities to observe an unmixed event,
 \begin{equation}
 P(\BzBzbar \rightarrow \BzBzbar) \propto 
 e^{-|\Dt|/\tauBz}(1 + \cos\Dm\Dt),
 \label{eq:trueP1}
  \end{equation} 
or a mixed event,
 \begin{equation}
 P(\BzBzbar \rightarrow \Bz\Bz\ {\rm or}\ \Bzbar\Bzbar)\! \propto \!
 e^{-|\Dt|/\tauBz}(1 - \cos\Dm\Dt).
 \label{eq:trueP2}
  \end{equation} 
Therefore, if we measure $\Dt$ and identify the $b$-quark flavor
of both $B$ mesons at their time of decay, we can extract 
\tauBz and \Dm.
In this analysis, one $B$ is reconstructed in the mode \btodstlnu, 
which has a measured  branching fraction of 
$(4.60 \pm 0.21)\%$~\cite{ref:PDG2002}.  
Although the neutrino cannot be detected, 
the requirement of a reconstructed \dsttodpi
decay and a
high-momentum lepton satisfying kinematic constraints consistent with the decay
\btodstlnu allows the isolation of a signal sample with 
(65--89)\% purity, depending on the \dz decay mode
and whether the lepton candidate is an electron or a muon.
The charges of the
final-state particles identify the meson as a \Bz or a \Bzbar.
The remaining charged particles in the event, which originate from the other 
$B$ (referred to as \btag), are used to identify, or ``tag", its 
flavor as a \Bz or a \Bzbar.
The time difference 
$\Dt = \tdstl - \ttag \approx \Dz/\beta\gamma c$
is determined from the separation $\Dz$ of the decay vertices 
for the \dstl candidate and the tagging $B$ along the boost direction.
The average separation is about 250~$\mu$m.

The oscillation frequency and the average lifetime
of the neutral $B$ meson are determined simultaneously with
an unbinned maximum-likelihood fit to the measured $\Dt$ distributions
of events that are classified as mixed and unmixed.
This is in contrast to most published measurements~\cite{ref:PDG2002,ref:OPAL} in
which only \tauBz is measured, or 
\Dm is measured with \tauBz fixed to the world average.
There are several reasons to measure the lifetime and oscillation 
frequency simultaneously.
The statistical precision of this measurement for both \tauBz and \Dm
is comparable  
to the uncertainty on the  world average; hence, it is appropriate to
measure both quantities rather than
fixing the lifetime to the world average.
Since mixed and unmixed events have different \Dt distributions,
the separation of mixed and unmixed events gives greater sensitivity to
the \Dt resolution function; as a result, the statistical uncertainty
of \tauBz is improved by approximately 15\%~\cite{vtxBs}.
Also, since $B^+B^-$ events do not mix, 
we can use the \Dt distributions for mixed and unmixed
events to help discriminate between \BzBzbar signal events 
and $B^+B^-$ background events in the lifetime and mixing measurement.

There are three main experimental complications that affect the $\Dt$ 
distributions given in Eqs.~\ref{eq:trueP1} and \ref{eq:trueP2}.
First, the tagging algorithm, which classifies events into 
categories $c$ depending on the source of the available
tagging information, incorrectly identifies the flavor of \btag with
a probability $w_c$ with a consequent reduction of the observed
amplitude for the mixing oscillation by a factor 
$(1-2w_c)$.
Second, the resolution for $\Dt$ is comparable to the 
lifetime and must be well understood.
The probability density functions (PDF's) for the unmixed ($+$) and mixed ($-$)
signal events can be expressed as the 
convolution of the underlying \Dttrue distribution for  
tagging category $c$, 
\begin{equation}
 {e^{-|\Dttrue|/\tauBz}\over 4\tauBz}
  [1\pm(1-2w_c)\cos\Dm\Dttrue]\,,
\end{equation}
with a resolution function 
that depends on a set of parameters determined from the data.
A final complication is that the sample of selected \btodstlnu
candidates includes several types of background for which the \Dt
distributions must be determined.

To characterize the backgrounds,
we select control samples of events enhanced
in each type of background
and determine the signal and the background probabilities for
each event in the signal samples and the background control samples
as described in Sec.~\ref{sec:evtsel}.
The measurement of \Dz and the determination of 
\Dt and the uncertainty on \Dt (\sigmaDt) for each event is discussed
in Sec.~\ref{sec:dectime}. 
The $b$-quark tagging algorithm is described in Sec.~\ref{sec:tagging}.
In Sec.~\ref{sec:fitmodel}, we describe the unbinned maximum-likelihood fit.
The physics model and \Dt resolution function used to describe the 
measured \Dt distribution for signal are given in Sec.~\ref{sec:sigmodel}.
A combination of Monte Carlo simulation and data samples are used to determine
the parameterization of the PDF's to describe the \Dt distribution for each type
of background, as described in Sec.~\ref{sec:bkgndmodel}.
The likelihood is maximized in a simultaneous fit to the signal and 
background control samples to extract 
the \Bz lifetime \tauBz,
the mixing frequency \Dm, 
the mistag probabilities $w_c$, 
the signal \Dt resolution parameters $\vec{q}_c$,
the background \Dt model parameters,
and the fraction of $B^+\rightarrow\dstlnu X$ decays in the signal sample.
The results of the fit are given in Sec.~\ref{sec:results}.
Cross-checks are described in Sec.~\ref{sec:validations} and 
systematic uncertainties are summarized in Sec.~\ref{sec:systematics}.

%%%%%%%%%%%%%%%%%%%%%%%%%%%%%%%%%%%%%%%%%%%%%

\section{The \babar detector}

\label{sec:detector}

The \babar detector is described in detail elsewhere~\cite{ref:babar}.
The momenta of charged particles are measured with a
combination of 
a five-layer silicon vertex tracker (SVT) and
a 40-layer drift chamber (DCH)
in a 1.5-T solenoidal magnetic field.
A detector of internally-reflected Cherenkov radiation (DIRC) is used
for charged particle identification.
Kaons are identified with a neural network based on the likelihood ratios 
calculated from d$E$/d$x$ measurements in the SVT and DCH, and from the observed
pattern of Cherenkov light in the DIRC.
A finely-segmented CsI(Tl) electromagnetic calorimeter (EMC) is used to detect photons
and neutral hadrons, and to identify electrons.
Electron candidates are required to have a 
ratio of EMC energy to track momentum, 
an EMC cluster shape,
DCH d$E$/d$x$, 
and DIRC Cherenkov angle all consistent with the electron hypothesis.
The instrumented flux return (IFR) contains resistive plate
chambers for muon and long-lived neutral hadron identification.
Muon candidates are required to have IFR hits
located along the extrapolated DCH track, 
an IFR penetration length,
and an energy deposit in the EMC consistent with
the muon hypothesis.

%%%%%%%%%%%%%%%%%%%%%%%%%%%%%%%%%%%%%%%%%%%%%

\section{Data samples}

\label{sec:samples}

The data used in this analysis were recorded with the \babar
detector at the PEP-II storage ring
in the period October 1999 to December 2000.
The total integrated luminosity of the data set is 20.6~fb$^{-1}$ 
collected at the \FourS resonance and 2.6~fb$^{-1}$ collected 
about 40~\MeV below
the \FourS (off-resonance data).
The corresponding number of produced $B\overline B$ pairs is 23 million.

Samples of Monte-Carlo simulated $B\overline B$  and $c\overline c$
events, generated with a GEANT3~\cite{ref:GEANT3} detector simulation,
are analyzed through the same analysis chain as the data to check
for biases in the extracted physics parameters and are also used to
develop  models for describing physics and detector resolution
effects. However, the values of the parameters used in these models
are determined with 
data. The equivalent luminosity of this simulated sample is
approximately equal to that of the data for $B\overline B$ events
and about half that of data for $c\overline c$ events. 
In addition, we generate signal Monte Carlo samples in which one
neutral $B$ meson in every event decays to \dstlnu, with \dsttodpi,
and the other neutral $B$ meson decays to any final
state~\cite{ref:evtgen}. The \dz then 
decays to one of the four final states reconstructed in this analysis
(described in the next section). The equivalent luminosity of the
simulated signal samples is between 2 and 8 times that of the
data, depending on the \dz decay mode. 

%%%%%%%%%%%%%%%%%%%%%%%%%%%%%%%%%%%%%%%%%%%%%

\section{Event selection and characterization}

\label{sec:evtsel}

We select events containing a fully-reconstructed 
\dstm and an identified oppositely-charged electron or muon. 
This $\dstl$ pair is then required to pass kinematic cuts that enhance the
contribution of semileptonic \btodstlnu decays.  
In  addition to the signal sample, we select several control samples
that are used to characterize the main sources of background.

We define the following classification of the sources of signal and
background that we expect to contribute to this sample.
The nomenclature shown in italics will be used throughout this paper
to define signal and all possible types of background. 
Events are classified according to the \dstm candidate reconstruction
status and the source of the lepton candidate.
\begin{enumerate}
\item Events with a correctly reconstructed \dstm candidate:
  \begin{enumerate}
  \item Events that originate from $B\overline B$ events:
    \begin{enumerate}
    \item Events with a correctly identified lepton candidate:
      \begin{enumerate}
      \item {\it Signal} -- \btodstlnu(X) decays, where the \dstm and 
       lepton originate from a common point. 
       $(X)$ indicates
       the possibility of one or more pions or photons from the direct
       decay of the parent $B$ or from the decay of short-lived 
       intermediate resonances (radially- and orbitally-excited $D$ states).
      \item {\it Uncorrelated-lepton background} -- 
        events in which the lepton does not come from the primary
        $B$ decay that produced the \dstm:
        ($B\rightarrow \dstm X,\ {\rm other}\ {B}\rightarrow\ell^+ X$) or
        ($B\rightarrow \dstm X,\ X\rightarrow\ell^+ Y$). 
      \item {\it Charged $B$ background} -- 
        $B^+\rightarrow \dstm\ell^+\nu_\ell X$.
      \end{enumerate}
    \item {\it Fake-lepton background} -- events with a misidentified
        lepton candidate.
    \end{enumerate}
  \item {\it Continuum background} -- $c\overline c \rightarrow\dstm X$.
  \end{enumerate}
\item {\it Combinatorial-\dst background} -- events with a misreconstructed
  \dstm candidate.
\end{enumerate}

\subsection{Lepton candidates}

Lepton candidates are defined as tracks with momentum 
greater than 1.2~\GeVc in the \FourS rest frame. 
For the \dste samples, the electron candidate passes selection
criteria with a corresponding electron identification efficiency of
about 90\% and 
hadron misidentification less than 0.2\%.
For the \dstmu samples, the muon candidate passes selection
criteria with a corresponding muon identification efficiency of about 70\% and
hadron misidentification between 2\% and 3\%. The particle
identification criteria in \babar are described in detail
elsewhere~\cite{ref:PRD}. 
A sample enriched in fake-lepton background is also selected, where 
\dstl candidates are accepted if the 
lepton {\em fails}
both electron and muon selection criteria looser than those required
for lepton candidates.
This sample is used to determine the fraction and \Dt distribution of
the fake-lepton background.

\subsection{\dstm candidates}
\dstm candidates are selected in the decay mode \dsttodpi. 
The \Dzbar candidate is reconstructed in the modes
\kpi, \kpipipi, \kpipiz and \kspipi
The daughters of the \Dzbar decay are selected according to the following
definitions. 
$\pi^0$ candidates are reconstructed from two photons with energy
greater than 30~\MeV each, and an invariant mass between 119.2 and
150.0~\MeVcc and a total energy greater then 200~\MeV.
The mass of the photon pair is constrained to the $\pi^0$ mass and 
the photon pair is kept as a $\pi^0$ candidate if the \chisq
probability of the fit 
is greater than 1\%.
\ks candidates are reconstructed from a pair of charged particles with 
invariant mass within 15~\MeVcc of the \ks mass. 
The pair of tracks is retained as a \ks candidate if the \chisq
probability that the 
two tracks form a common vertex is greater than 1\%.
Charged kaon candidates satisfy loose kaon criteria~\cite{ref:PRD} for the
\kpi mode and tighter criteria for the \kpipipi and
\kpipiz modes. 
For the \kpipiz and \kspipi modes, a likelihood is
calculated as the square of the decay amplitude in the Dalitz plot for
the three-body candidate, based on measured amplitudes and phases
\cite{ref:E687}. The candidate is retained if the likelihood is
greater than 10\% of its maximum value across the Dalitz plot. 
This criterion rejects about 95\% (97\%) of uniform background
and has a signal efficiency of about 62\% (48\%) for the
\kpipiz (\kspipi) mode if the real signal is
described by the results in Ref.~\cite{ref:E687}.

\Dzbar candidates in the \kpi, \kpipipi, and
\kspipi modes (\kpipiz mode) are selected if
they have an invariant 
mass within 17~\MeVcc (34~\MeVcc) of the \dz mass.
The invariant mass of the daughters is constrained to the \dz mass and
the tracks 
are constrained to a common vertex in a simultaneous fit.  
The \Dzbar candidate is retained if the \chisq probability of the fit
is greater than 0.1\%. 

The low-momentum pion candidates for the \dsttodpi decay are selected
with total momentum less
than  450~\MeVc in the \FourS rest frame
and momentum transverse to the beamline greater than 50~\MeVc.
The momentum of the \dstm candidate in the \FourS rest frame 
is required to be between 0.5 and 2.5~\GeVc. The requirements on the
momenta of the low-momentum pion and \dstm candidates retain
essentially all signal events and 
reject higher momentum \dstm from continuum events.
\dstm candidates are rejected if $|\cos\theta^*_{\rm thrust}| \ge
0.85$, where $\theta^*_{\rm thrust}$ is the angle between the thrust
axis of the \dstl candidate and the thrust axis of the remaining
charged and neutral particles in the event. The distribution of 
$|\cos\theta^*_{\rm thrust}|$ is peaked at 1 for jet-like continuum
events and is flat for more spherical $B\overline B$ events.

\dstm candidates are retained if $\massdiff$ is less than 165~\MeVcc,
where $m(\dst)$ is the candidate $\Dzbar\pi^-$ mass calculated with 
the candidate $\Dzbar$ mass constrained to the true \dz mass, 
$m(\dz)$. Note that the \massdiff distribution has a kinematic
threshold at the mass of the $\pi^-$, and a peak at 145.5~\MeVcc
with a resolution of 1~\MeVcc or better.
We have retained the sideband of
the \massdiff distribution for studies of combinatorial-\dst background.

\subsection{\dstl candidates}

\dstl candidates are retained if the following criteria are met:
the \chisq probability of the fit of
the lepton, $\pi^-$, and \Dzbar candidates to a common vertex is
greater than 1\%;
the decay point of \btag is determined from at least two tracks;
the fit that determines the distance \Dz between the two $B$
decays along the beamline converges; the time between decays (\Dt)
calculated from \Dz is less than 18~ps; and the calculated error on
\Dt (\sigmaDt) is less than 1.8~ps.
See Sec.~\ref{sec:dectime} for details on the determination of the 
decay point of \btag and the calculation of \Dt and \sigmaDt.

We define two angular quantities for each \dstl candidate to classify
them into a sample 
enriched in \btodstlnu signal events, 
and a sample enriched in {\it uncorrelated-lepton} background 
events.
The first angle is $\theta_{\dst,\ell}$, the angle between the \dstm and
lepton candidate in the \FourS rest frame.  
The second is \thby, the inferred angle between the direction of the
\Bz and the vector sum of the \dstm and lepton candidate momenta, 
calculated in the \FourS rest
frame. 
Since we do not know the direction of the \Bz, we calculate the cosine of \thby from
the following equation, in which we assume that the only $B$ decay particle missed in
the reconstruction is a massless neutrino:
 \begin{equation}
 \costhby \equiv {-(m^2_{\Bz} + m^2_{\dst\ell} - 2 E_B E_{\dst\ell}) \over 
              2 |\vec{p}_B| |\vec{p}_{\dst\ell}|}.
 \label{eq:costhby}
 \end{equation} 
All quantities in Eq.~\ref{eq:costhby} are defined in the \FourS rest frame.
The energy and the magnitude of the momentum of the $B$ are 
calculated from the 
$e^+e^-$ center-of-mass energy and the \Bz mass.
For true \btodstlnu events, \costhby lies in the physical
region $[-1, +1]$, except for detector resolution effects.
Backgrounds lie inside and outside the range $[-1, +1]$.
We also calculate the same angle with the lepton momentum direction 
reflected through the origin in the \FourS rest frame: \thbyfl.
This angle is used to select samples enriched in uncorrelated-lepton
background. 

A sample enhanced in \btodstlnu signal events 
(called the {\it opposite-side} sample)
is composed of \dstl candidates with $\cos\theta_{\dst\ell}<0$ and 
$|\costhby|<1.1$.
Samples are defined for lepton candidates that satisfy the criteria
for an electron, a 
muon and a fake-lepton.
The first two samples are the signal samples, and the latter is the
{\it fake-lepton} control sample.

An additional background control sample, representative of the 
uncorrelated-lepton background and called the {\it same-side} sample,
is composed of  
\dstl candidates satisfying $\cos\theta_{\dst\ell}\geq 0$ and
$|\costhbyfl|<1.1$.  
We use \costhbyfl rather than \costhby because, in 
Monte Carlo simulation,
the distribution of \costhbyfl in this
control sample is similar to the distribution of \costhby for
uncorrelated-lepton background in the signal sample, whereas the
distribution of \costhby in the background control sample is
systematically different.

\subsection{Signal and background subsamples}
Approximately $68,000$ candidates pass the above selection criteria. 
These candidates are distributed over two signal samples 
and ten background control samples
defined by the following characteristics:
\begin{enumerate}
\item whether the data were recorded on or off the \FourS resonance (two
choices); 
\item whether the candidate lepton is {\it same-side} or {\it opposite-side}
to the \dstm
candidate (two choices); 
\item whether the lepton candidate passes the criteria for an electron, a
muon, or a fake lepton (three choices). 
\end{enumerate}
The signal samples are the electron and muon
samples in the opposite-side, on-resonance data.

\begin{figure}[!htb]
\begin{center}
\includegraphics[width=2.5in]{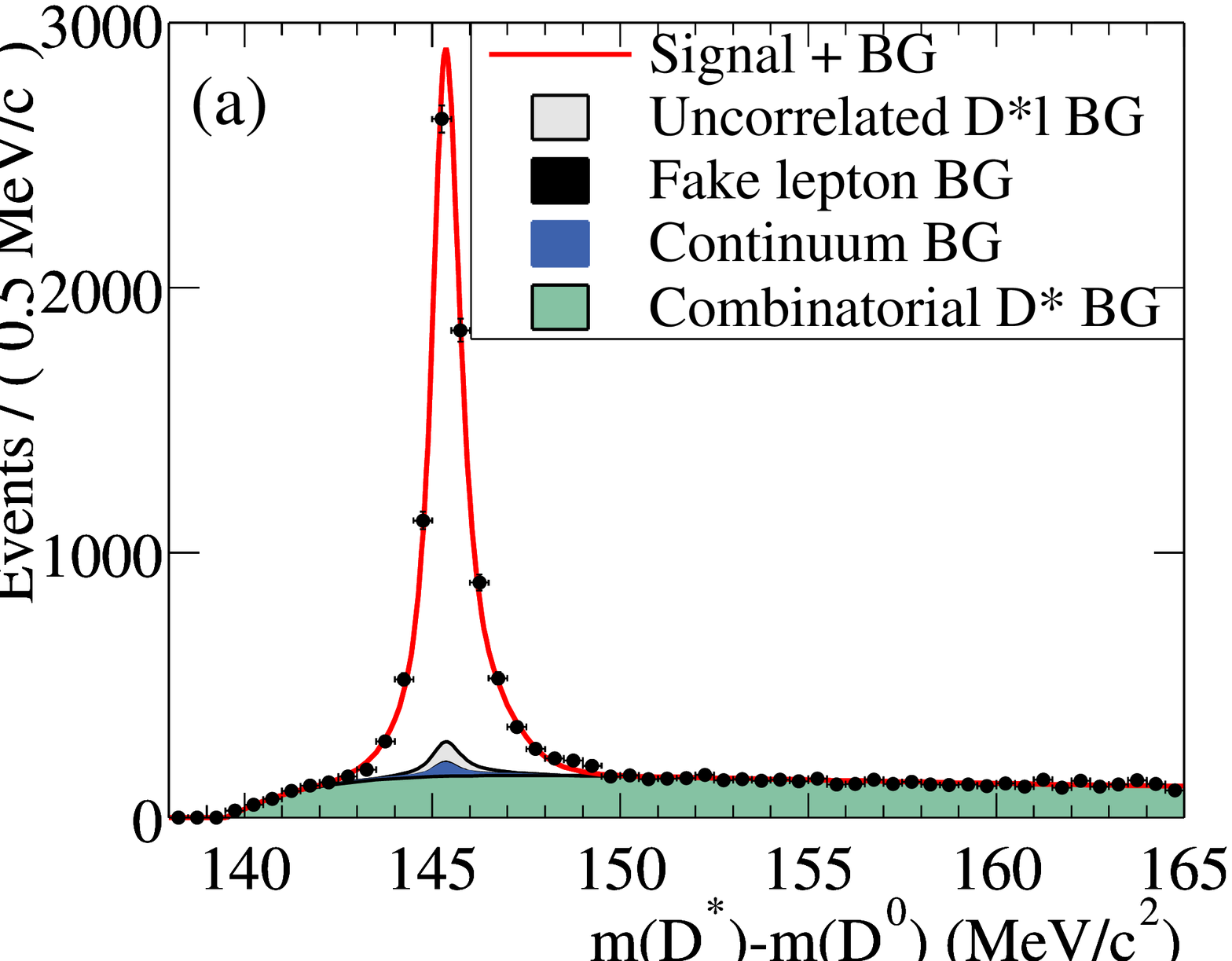}
\includegraphics[width=2.5in]{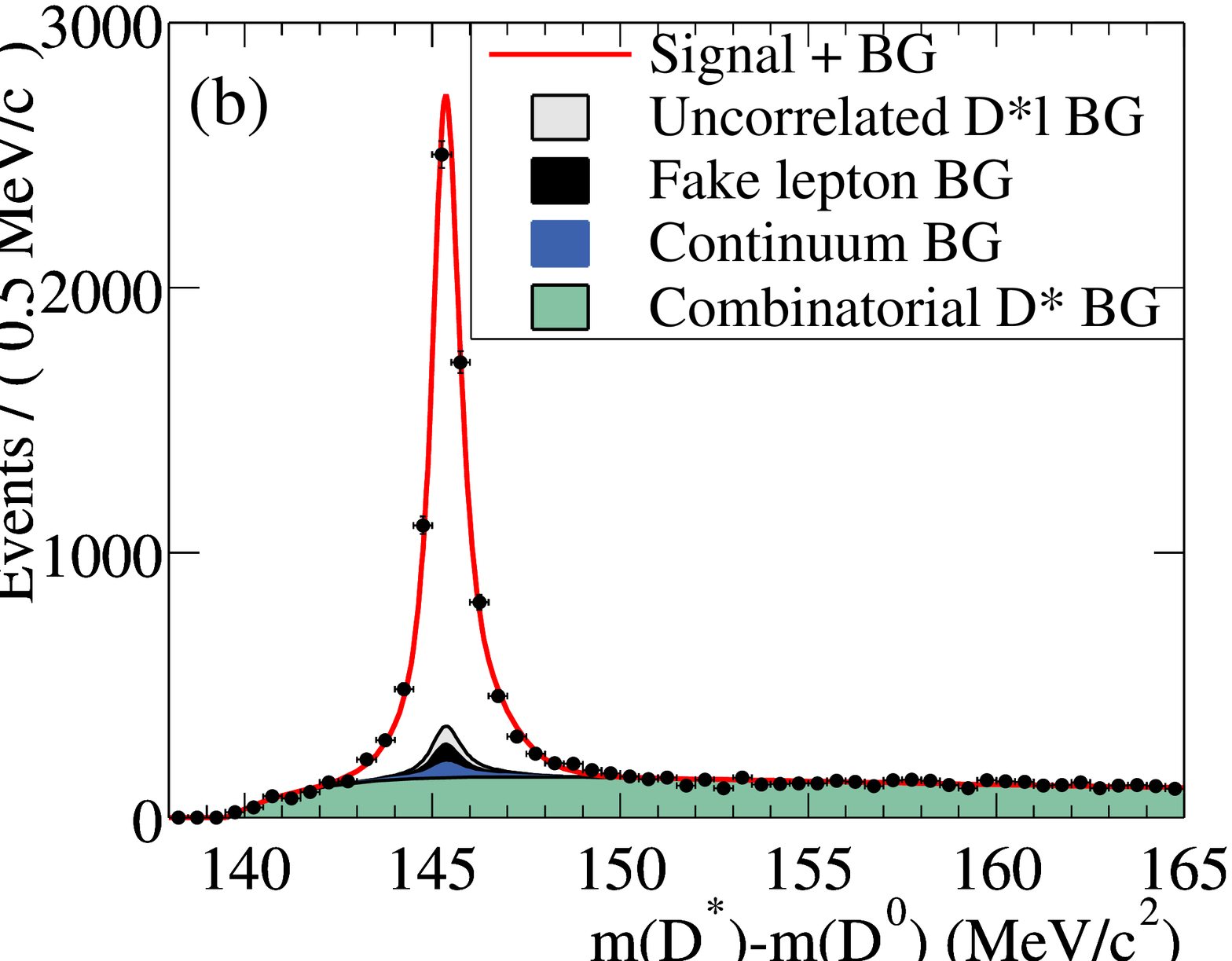}
\caption{\massdiff distribution for events passing all selection criteria for
\btodstlnu, with (a) an electron or (b) a muon candidate.  
The points correspond to the data.
The curve is the result of a simultaneous unbinned maximum likelihood fit to
this sample of events and a number of background control samples.
The shaded distributions correspond to the four types of background (BG)
described in the text. The charged $B$ background is not shown separately.}
\label{fig:massdifsig}
\end{center}
\end{figure}

\begin{figure}[!htb]
\begin{center}
\includegraphics[width=1.68in]{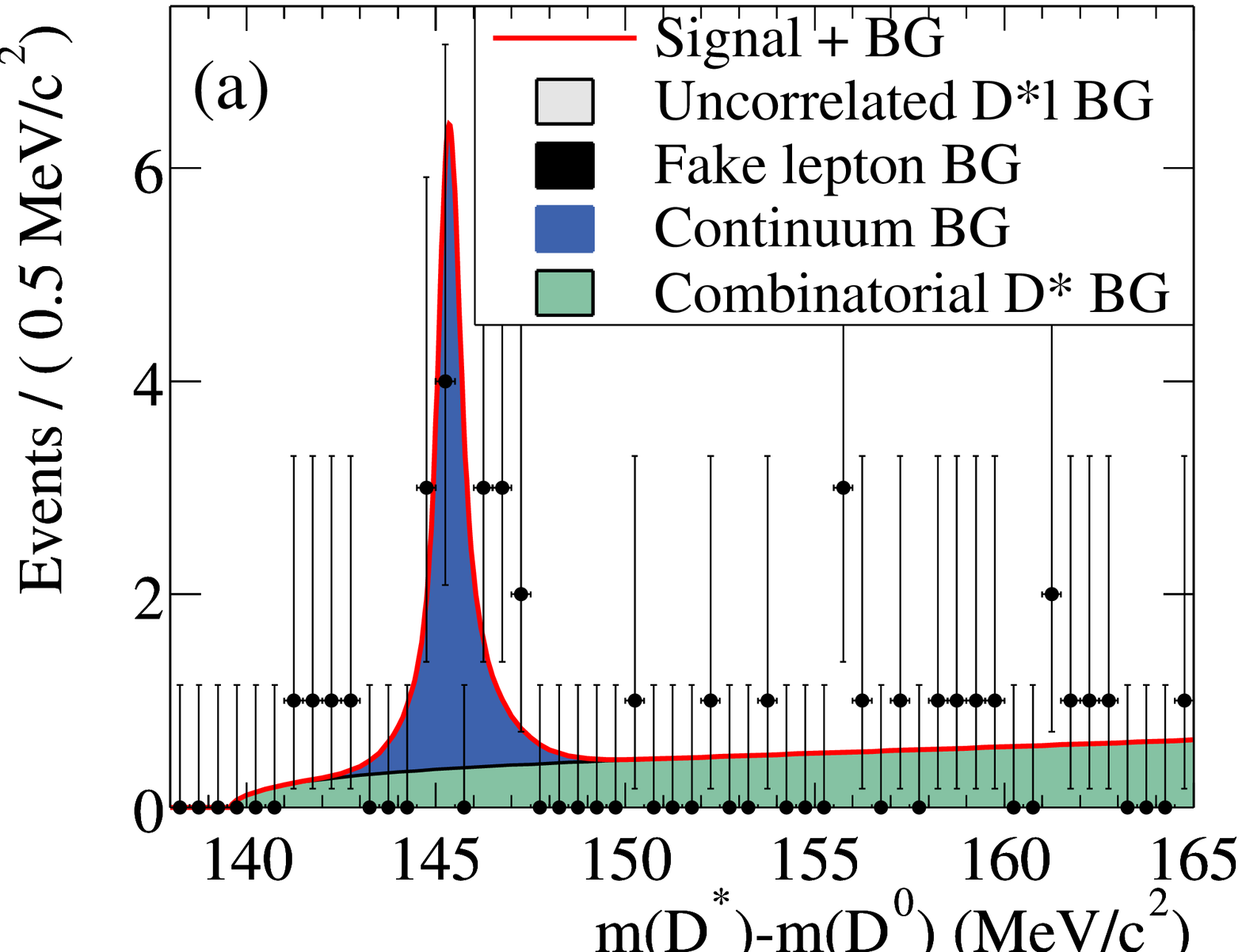}
\includegraphics[width=1.68in]{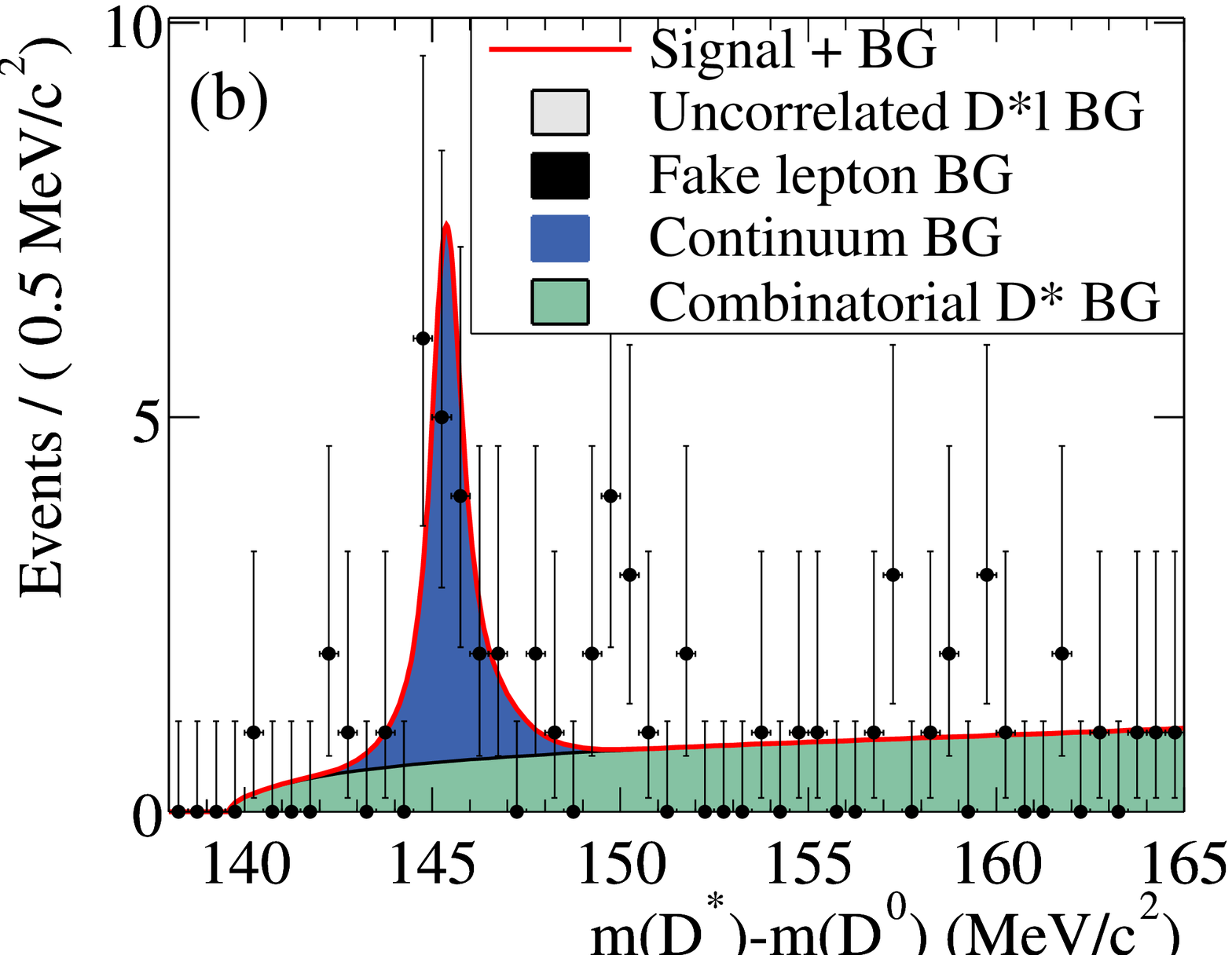}

\vspace{0.2cm}
\includegraphics[width=1.68in]{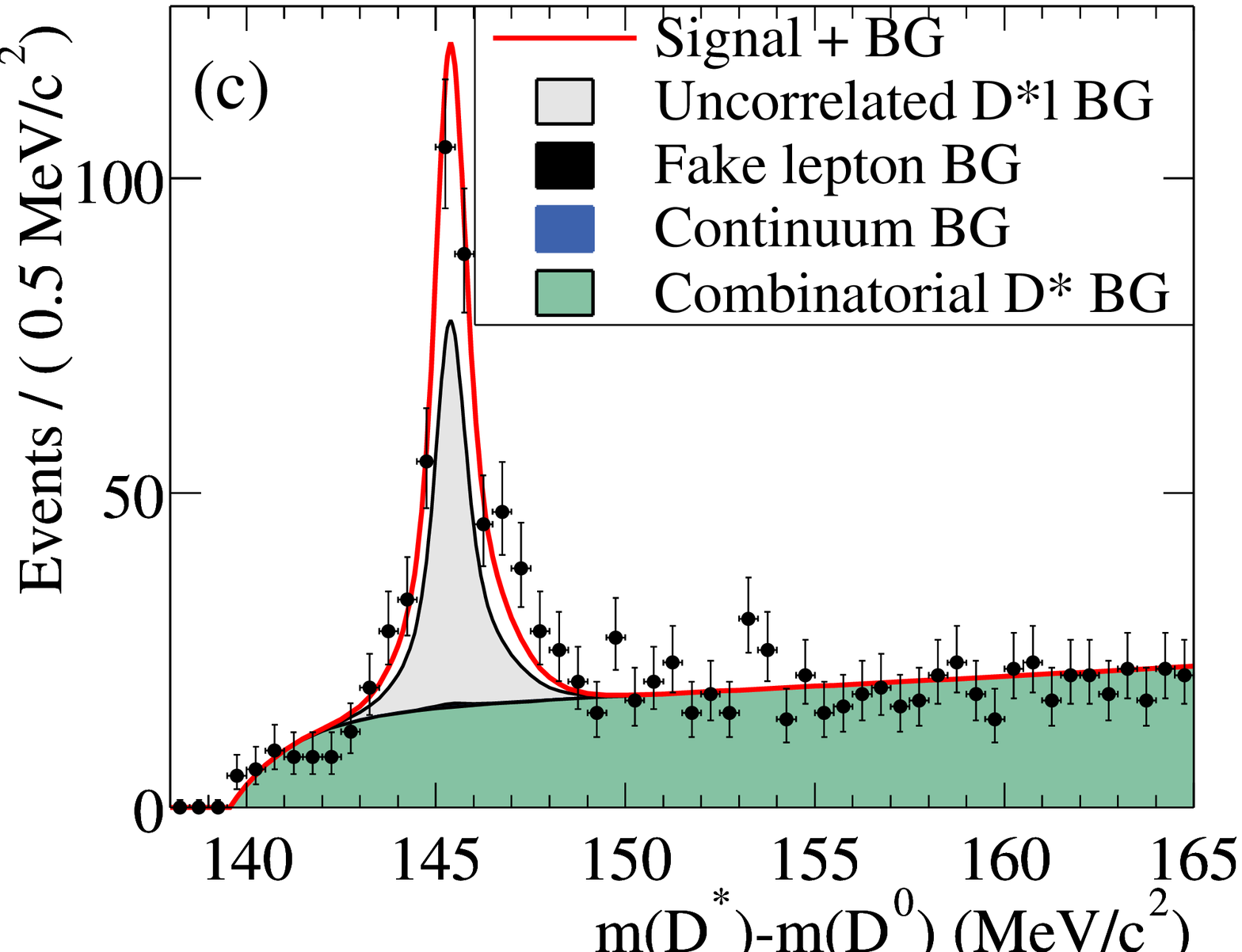}
\includegraphics[width=1.68in]{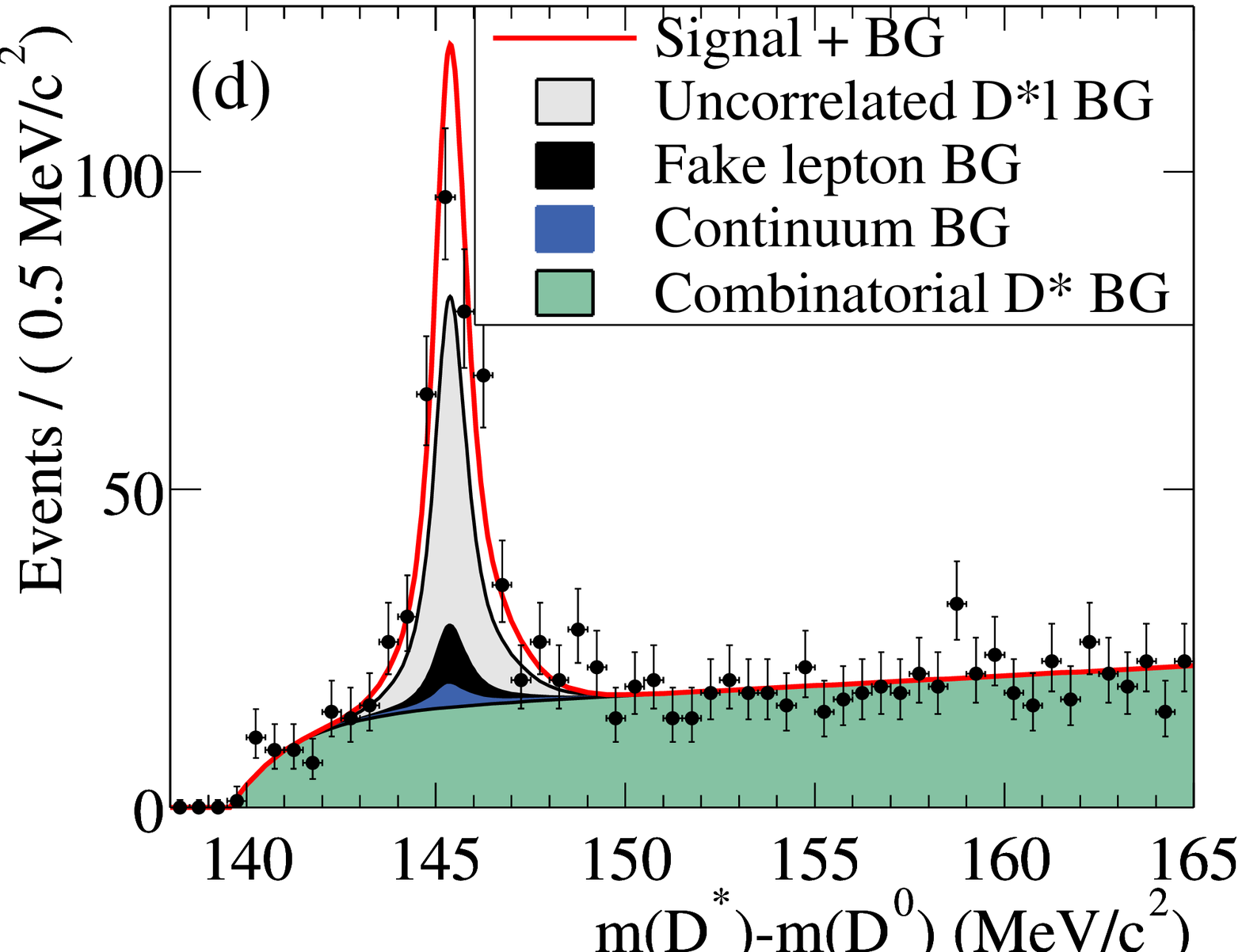}

\vspace{0.2cm}
\includegraphics[width=1.68in]{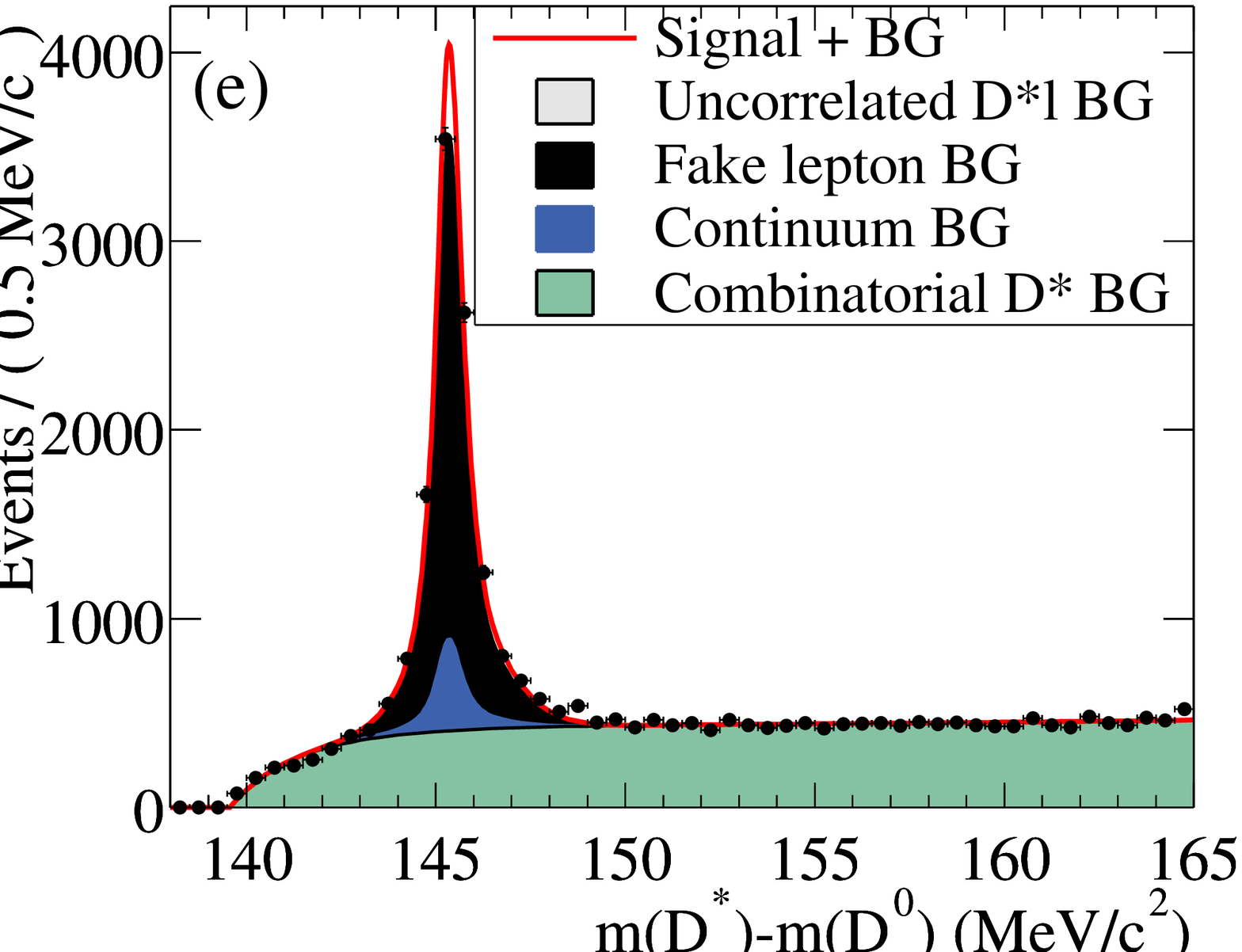}
\caption{\massdiff distribution for events passing all selection
criteria in background control samples:
opposite-side (a) \dste and (b) \dstmu candidates in
off-resonance data; 
same-side (c) \dste and (d) \dstmu\
candidates in on-resonance data; 
(e) opposite-side \dstm--fake-lepton candidates in
on-resonance data.
The points correspond to the data.
The curve is the result of a simultaneous unbinned maximum likelihood fit to
this sample of events, the signal sample, and a number of other background
control samples. 
The shaded distributions correspond to the four types of background 
described in the text. The charged $B$ background is not shown separately.}
\label{fig:massdifctrl}
\end{center}
\end{figure}

The combinatorial-\dst background 
can be distinguished from events with a real \dstm in a plot 
of the mass difference \massdiff.
The \massdiff distributions for the samples of signal events 
(opposite-side \dste and \dstmu candidates in on-resonance data) 
are shown as data points in
Fig.~\ref{fig:massdifsig} for (a) electron candidates
and (b) muon candidates. The contributions of the three types of
background that contain a real \dstm 
(continuum, uncorrelated-lepton, and fake-lepton, together
called the {\it peaking
background}), except for the charged $B$ background, are also shown in
the plots. 
The \massdiff distributions for five 
background control samples used for determining the background levels in
the signal sample
are shown as data points in
Fig.~\ref{fig:massdifctrl}: 
opposite-side (a) \dste and (b) \dstmu candidates in
off-resonance data; same-side (c) \dste and (d) \dstmu
candidates in on-resonance data; 
(e) opposite-side \dstm--fake-lepton candidates in
on-resonance data. The remaining five background control samples
are useful for determining the background levels in the first five
control samples.

Each of the 12 samples described above 
are further divided into 30 subsamples
according to the following characteristics that affect the \massdiff or \Dt\
distributions.
\begin{enumerate}
\item The $\pi^-$ from the \dstm decay reconstructed in the SVT only, or in the SVT
and DCH (two choices):  The \massdiff resolution is worse when the $\pi^-$ is
reconstructed only in the SVT.
\item The \Dzbar candidate reconstructed in the mode \kpi or \kpipiz or
(\kpipipi or \kspipi) (three choices): The level of contamination
from combinatorial-\dst background and  the \massdiff resolution depend on
the \Dzbar decay mode. 
\item The $b$-tagging information used for the other $B$ (five choices; see Sec.~\ref{sec:tagging}):
The level of contamination from each type of background and the \Dt resolution parameters 
depend on the tagging information.
\end{enumerate} 
This allows subdivision into 360 samples.
In the unbinned maximum likelihood fits to the \massdiff and (\Dt, \sigmaDt)
distributions, individual fit parameters are shared among different
sets of subsamples 
based on physics motivation and observations from the data.

We fit the \massdiff distributions to determine signal and background
probabilities for each of the 360 subsamples.
The peak due to real \dstm candidates is modeled by the sum of two Gaussian
distributions; 
the mean and variance of both the Gaussian distributions, as well as
the relative 
normalization of the two Gaussians, are free parameters in the fit. 
We model the shape
of the combinatorial-\dst background with the function
 \begin{equation}
 {1\over N} \left[ 1-\exp\left(-{\dm - m_{\pi^-} \over c_1}\right)\right] 
  \left({\dm\over m_{\pi^-} }\right)^{c_2},
 \label{eq:combback}
 \end{equation} 
where $\dm \equiv \massdiff$, $N$ is a normalization constant,
$m_{\pi^-}$ is the mass of the $\pi^-$, and $c_1$ and $c_2$ are free
parameters in the fit. An initial fit is performed to determine the shape
parameters describing the peak and combinatorial-\dst background. 
Separate values of the five parameters describing the shape of the
peak are used for the six subsamples defined by 
(1)
 whether the $\pi^-$
candidate is tracked in the SVT 
only or in both the SVT and DCH (two choices), and 
(2) 
the three types of \Dzbar decay modes.
Each of these six groups that use separate peak parameters is further
subdivided 
into twelve subgroups that each uses a different set of the 
two combinatorial-\dst shape parameters but the same set of peak parameters.
Ten of these twelve subgroups are defined by 
the five tagging
categories for the large signal samples and for the 
fake-lepton control samples, in on-resonance data.
The other two subgroups are defined as
same-side, on-resonance samples and 
all off-resonance samples.

Once the peak and combinatorial-\dst  shape parameters have been
determined, we fix the shape parameters and determine the peak and
combinatorial-\dst yields in each of the 360 subsamples with an
unbinned extended maximum-likelihood fit. 

The total peak yields in the
signal sample and each background control sample are then used 
to determine the amount of true signal and each type of 
peaking background in the 
\massdiff peak of each sample as follows.
\begin{enumerate}
\item
{\em Continuum background  --} \ \
For each subsample in on-resonance data, the peak yield of the
corresponding subsample in off-resonance data is scaled by the
relative integrated luminosity for on- and off-resonance data,
to determine the continuum-background yields in
on-resonance data.
\item
{\em Fake-lepton background --} \ \ 
Particle identification efficiencies and misidentification probabilities 
for the electron, muon, and fake-lepton selection criteria
are measured in separate data samples as a
function of laboratory momentum, polar angle, and azimuthal angle, for 
true electrons, muons, pions, kaons, and protons.  
$\Bz\Bzbar$ and $B^+B^-$ Monte Carlo simulations are used to determine the 
measured laboratory momentum,
polar angle, and azimuthal angle distributions for true 
electrons, muons, pions, kaons
and protons that pass all  selection criteria for \dstl candidates,
except the lepton (or fake-lepton) identification criteria. 
These distributions are combined with the measured particle identification
efficiencies and misidentification probabilities
to determine the momentum- and
angle-weighted probabilities for a true lepton or true hadron to pass
the criteria for 
a lepton or a fake lepton in each of the \dstl signal and background
control samples. 
We then use these efficiencies and misidentification probabilities, 
and the observed
number of electron, muon and fake-lepton candidates in each subsample
in data, after removing 
the continuum background contribution, to determine the
number of 
true leptons and fake leptons (hadrons) in each control sample.
\item
{\em Uncorrelated-lepton background  --}\ \ 
The relative efficiencies for signal and uncorrelated-lepton events to
pass the criteria for same-side and opposite-side samples are
calculated from Monte Carlo simulation. 
These efficiencies and the \massdiff peak yields, after removing the
continuum and fake-lepton background contributions, are used to
determine the number of uncorrelated-lepton events in each subsample.
\end{enumerate}

The peak yields and continuum, fake-lepton, and uncorrelated-lepton
fractions of 
the peak yield, as well as the combinatorial-\dst fraction of all events in
a \massdiff 
signal window, are shown in Table~\ref{tab:yields} 
for the signal and background control samples in on-resonance data.  
The peak yields include the peaking backgrounds. 
The signal window is defined as (143--148)~\MeVcc for the calculation of
the combinatorial-\dst background fractions.
Table~\ref{tab:sampleyields} shows the peak yields and the
combinatorial-\dst background fractions for different divisions of
the signal sample (opposite-side lepton candidates in on-resonance
data). This table demonstrates that the background levels vary
significantly among
subgroups of the signal sample.

\begin{table*}[!htb]
\caption{Peak yields and the fraction of them that are due to
continuum, fake-lepton, and
uncorrelated-lepton events.
Also shown is the combinatorial-\dst 
fraction of total events in a \massdiff
signal window for the signal and background control samples 
in on-resonance data.  Peak yields
include the peaking backgrounds.  
The signal window for combinatorial-\dst background
fractions is defined as (143--148)~\MeVcc.
$e$, $\mu$, and fake indicate the type of lepton candidate: electron, muon 
or fake-lepton.}
\begin{center}
\begin{tabular*}{0.75\textwidth}{@{\extracolsep{\fill}}lccccc}
\hline\hline
Category & Peak Yield & $f_{\mathrm{cont}}(\%)$ & $f_{\mathrm{fake}}(\%)$
& $f_{\mathrm{uncor}}(\%)$ & $f_{\mathrm{comb}}(\%)$ \\
\hline
{\it opposite-side} & \multicolumn{5}{c}{} \\
\qquad $e$ & $ 7008\pm 91$ & $ 1.5\pm 0.4$ & $ 0.168\pm 0.004$ & $ 3.1\pm 0.4$ & $ 17.9\pm 0.2$ \\
\qquad $\mu$ & $ 6569\pm 88$ & $ 2.3\pm 0.6$ & $ 2.67\pm 0.07$ & $ 2.9\pm 0.5$ & $ 18.4\pm 0.3$ \\
\qquad fake & $ 8770\pm 108$ & $ 12.8\pm 1.3$ & $ 72.4\pm 1.8$ & $ 0.7\pm 1.6$ & $ 31.4\pm 0.2$ \\
\\
{\it same-side} & \multicolumn{5}{c}{} \\
\qquad $e$ & $ 306\pm 21$ & \;\;$<5.9^*$ & $ 0.53\pm 0.04$ & $ 56.9\pm 7.0$ & $ 34.0\pm 1.3$ \\
\qquad $\mu$ & $ 299\pm 20$ & $ 5.1\pm 3.6$ & $ 8.9\pm 0.6$ & $ 48.9\pm 8.0$ & $ 34.4\pm 1.3$ \\
\qquad fake & $ 1350\pm 45$ & $ 20.4\pm 4.1$ & $ 74.4\pm 5.4$ & $ 3.6\pm 7.8$ & $ 42.6\pm 0.6$ \\
\hline\hline
\multicolumn{5}{c}{} & \multicolumn{1}{c}{*90\% C.L.} \\
\end{tabular*}
\end{center}
\label{tab:yields}
\end{table*}

\begin{table}
\caption{Peak yields and the combinatorial-\dst background fraction of total
events in a \massdiff signal window for different divisions of
the signal sample (opposite-side lepton candidates in on-resonance 
data). In the first block, the signal sample is divided according to the
reconstruction status of the $\pi^-$ from the \dstm decay; the
second block by the \Dzbar decay mode; and the third block by the
$b$-tagging information (see Sec.~\ref{sec:tagging}).
The signal window for combinatorial-\dst background
fractions is defined as (143--148)~\MeVcc.}
\begin{center}
\begin{tabular*}{0.48\textwidth}{@{\extracolsep{\fill}}lcc}
\hline\hline
Category & Peak Yield & $f_{\mathrm{comb}}(\%)$ \\
\hline
$e$ & & \\
\qquad SVT only & $5427\pm 81$ & $19.5\pm 0.3$ \\
\qquad DCH \& SVT& $1581\pm 41$ & $11.8\pm 0.4$ \\
$\mu$ & & \\
\qquad SVT only & $5053\pm 78$ & $20.3\pm 0.3$ \\
\qquad DCH \& SVT & $1517\pm 41$ & $11.1\pm 0.4$ \\
\\
$e$ & & \\
\qquad $K\pi$ & $ 2623\pm 53$ & $ 7.0\pm 0.3$ \\
\qquad $K\pi\pi\pi\;\&\;\ks\pi\pi$ & $ 2219\pm 54$ & $ 28.6\pm 0.5$ \\
\qquad $K\pi\pi^0$ & $ 2166\pm 51$ & $ 16.9\pm 0.5$ \\
$\mu$ & & \\
\qquad $K\pi$ & $ 2491\pm 52$ & $ 7.4\pm 0.3$ \\
\qquad $K\pi\pi\pi\;\&\;\ks\pi\pi$ & $ 1939\pm 51$ & $ 30.9\pm 0.5$ \\
\qquad $K\pi\pi^0$ & $ 2139\pm 50$ & $ 16.1\pm 0.4$ \\
\\
$e$ & & \\
\qquad \lepton  & $ 783\pm 29$ & $ 8.2\pm 0.6$ \\
\qquad \kaon    & $ 2565\pm 55$ &$ 17.9\pm 0.4$ \\
\qquad \ntone   & $ 630\pm 27$ & $ 14.3\pm 0.8$ \\
\qquad \nttwo   & $ 921\pm 33$ & $ 20.9\pm 0.7$ \\
\qquad \ntthree & $ 2108\pm 51$ & $ 20.7\pm 0.5$ \\
$\mu$ & & \\
\qquad \lepton  & $ 746\pm 28$ & $ 8.3\pm 0.6$ \\
\qquad \kaon    & $ 2393\pm 53$ & $ 18.6\pm 0.4$ \\
\qquad \ntone   & $ 545\pm 25$ & $15.1\pm 0.8$ \\
\qquad \nttwo   & $ 958\pm 34$ & $19.4\pm 0.7$ \\
\qquad \ntthree & $ 1928\pm 49$ & $21.8\pm 0.5$ \\
\hline\hline
\end{tabular*}
\end{center}
\label{tab:sampleyields}
\end{table}

Finally, we use the calculated fractions and fitted shapes of the
background sources in each control sample to
estimate the probability of each candidate to be signal or each type of
background (combinatorial-\dst, continuum, fake-lepton, or uncorrelated-lepton)
when we fit the (\Dt, \sigmaDt) distribution to determine the lifetime
and mixing parameters.   
We take advantage of the fact that charged and neutral $B$ decays have
different decay-time distributions (because the charged $B$ does not mix) to 
determine the fraction of charged $B$ background events in the fit to 
(\Dt, \sigmaDt).

%%%%%%%%%%%%%%%%%%%%%%%%%%%%%%%%%%%%%%%%%%%%%

\section{Decay-time measurement}

\label{sec:dectime}

The decay-time difference \Dt between $B$ decays is determined from
the measured 
separation $\Dz \equiv \zdstl - \ztag$ along the $z$ axis 
between the \dstl vertex position
(\zdstl) and the flavor-tagging decay \btag vertex position (\ztag). 
This measured \Dz is converted into \Dt with the use of the \FourS boost,
determined from the beam energies for each run.
Since we cannot reconstruct the direction of the $B$ meson for each event, 
we use the approximation $\Dt \approx \Dz/(\beta\gamma c)$. Without
detector resolution effects, this approximation has a bias that depends
on the sum of the proper decay times ($t_1+t_2$) of the two $B$ mesons
and their 
direction in the \FourS rest frame~\cite{ref:PRDboost}. Neither of these
quantities can be measured 
because the \FourS production point is not known and the momentum of the
$B$ is not fully reconstructed due to a missing neutrino. After
integrating over $t_1+t_2$ and the $B$ meson direction,
the mean and RMS of the bias are 0 and 0.2~ps, respectively.

The momentum and position vectors of the \Dzbar, $\pi^-$, and lepton
candidates, and the  
average position of the $e^+e^-$ interaction point (called the beam
spot) in the plane 
transverse to the beam  are used in a constrained fit to determine the
position of the \dstl vertex. 
The beam-spot constraint is about 100~$\mu$m in the horizontal
direction and 30~$\mu$m in the vertical direction, corresponding to
the RMS size of 
the beam in the horizontal direction and the approximate transverse
flight path of the 
$B$ in the vertical direction.
The beam-spot constraint improves the resolution on \zdstl by about
20\% in Monte Carlo simulation; the RMS spread on the difference
between the measured and true position 
of the \dstl vertex is about 70~$\mu$m (0.4~ps).

We determine the position of the \btag vertex from all  tracks
in the  event except the daughters of the \dstl candidate, using 
$\ks\rightarrow \pi^+ \pi^-$ and
$\Lambda\rightarrow p \pi^-$ 
candidates in place of their daughter tracks, and excluding
tracks that are consistent with photon conversions. 
The same beam-spot constraint applied to the \bdstl vertex is also
applied to the \btag vertex.
To reduce the influence of charm decay products, which bias the
determination of the vertex position, 
tracks with a large contribution to the \chisq of
the vertex fit are iteratively removed until no track has a \chisq
contribution greater than 6
or only one track remains.  
The RMS spread on the difference between the measured and true position
of the \btag vertex in Monte Carlo simulation is about 160~$\mu$m (1.0~ps).
Therefore, the \Dt resolution is dominated by the $z$ resolution of
the tag vertex position. 

For each event,
we calculate the uncertainty on \Dz ($\sigma_{\Dz}$) due to
uncertainties on the track 
parameters from the SVT and DCH hit resolution and multiple
scattering, our knowledge of the beam-spot 
size, and the average $B$ flight length in the vertical direction.
The calculated uncertainty does not account for errors in 
pattern recognition in tracking, 
errors in associating tracks with the $B$ vertices,
the effects of misalignment within and between the tracking devices,
or the error on the approximation we use to calculate \Dt from \Dz.
The calculated uncertainties will also be incorrect if 
our assumptions for the amount
of material in the tracking detectors or the beam-spot size or 
position are inaccurate.
We use parameters in the \Dt resolution model, measured with data, to
account for 
uncertainties and biases introduced by these effects.

%%%%%%%%%%%%%%%%%%%%%%%%%%%%%%%%%%%%%%%%%%%%%

\section{Flavor tagging}

\label{sec:tagging}

All tracks in the event, except the daughter tracks of the \dstl candidate,
are used to determine whether the \btag decayed as a \Bz or a \Bzbar. 
This is called flavor tagging.
We use five different types of flavor tag, or tagging categories, in
this analysis. The first two tagging categories rely on the presence
of a prompt lepton, or one or more charged kaons, in the event. 
The other three categories exploit a variety of inputs with a
neural-network algorithm. The tagging algorithms are described briefly
in this section; see Ref.~\cite{ref:PRDtag} for more details.

Events are assigned a \lepton tag if they contain an identified lepton
with momentum in the \FourS rest frame greater than 1.0 or 1.1~\GeVc
for electrons and muons, respectively, thereby selecting mostly
primary leptons from the decay of the $b$ quark.
If the sum of charges of all identified kaons is nonzero, the event
is assigned a \kaon tag. The final three tagging categories 
are based on the output of a neural network that uses as inputs
the momentum and charge of the track with the maximum center-of-mass
momentum, the number of tracks with significant impact parameters with
respect to the interaction point, and the outputs of three other
neural networks, trained to identify primary leptons, kaons, and low
momentum pions from \dst decays. Depending on the output of the main
neural network, events are assigned to an \ntone (most
certain), \nttwo, or \ntthree (least certain) tagging category. 
About 30\% of events are in the \ntthree category, which has a mistag
rate close to 50\%.  
Therefore, these events are not sensitive to the mixing 
frequency, but they increase the sensitivity to the \Bz lifetime.

Tagging categories are mutually exclusive due to the hierarchical use
of the tags. Events with a \lepton tag and no conflicting \kaon tag
are assigned to the \lepton category.
If no \lepton tag exists, but the event has a \kaon tag, it is
assigned to the \kaon category. Otherwise events are assigned to
corresponding neural network categories.  
The mistag rates are free parameters in the final fit. The final
results are shown in Table~\ref{tab:result-signal} in
Sec.~\ref{sec:results}. 

%%%%%%%%%%%%%%%%%%%%%%%%%%%%%%%%%%%%%%%%%%%%%

\section{Fit method}

\label{sec:fitmodel}

We perform an unbinned fit simultaneously to events in each of the 12
signal and control samples 
(on or off resonance, opposite- or same-side lepton, 
electron or muon or fake lepton --
indexed by $s$) that are further
subdivided into 30 subsamples 
(tagging category, $D^0$ decay mode, with or without DCH hits for the 
 pion from the \dst decay --
indexed by $c$).
We maximize the likelihood
\begin{equation}
{\cal L} = \prod_{s=1}^{12}\, \prod_{c=1}^{30}\,\prod_{k=1}^{N(s,c)}\,
P_{s,c}(\dm_{k},\vec{x}_{k}\,;\vec{\eta}) \; ,
\label{eq:likelihood}
\end{equation}
where $k$ indexes the $N(s,c)$ events $\vec{x}_{k}$ in each of the 360
subsamples. 
The probability $P_{s,c}(\dm_{k},\vec{x}_{k}\,;\vec{\eta})$ of observing an event 
$(\dm_k,\vec{x}_k)$, where $\vec{x}_k = (\Dt,\sigmaDt,g)$, 
is calculated as a function of the parameters 
\begin{equation}
\begin{split}
\vec{\eta} = & (f_{s,c}^{\,\mathrm{comb}}, {\vec{p}}_{s,c}^{\,\mathrm{comb}},
{\vec{p}}_{c}^{\,\mathrm{peak}}, {\vec{q}}_{s,c}^{\,\mathrm{comb}},
f_{s,c,1}^{\,\mathrm{pkg}}, f_{s,c,2}^{\,\mathrm{pkg}},
f_{s,c,3}^{\,\mathrm{pkg}}, \\
& \qquad  
\fBp, {\vec{q}}_{s,c,1}^{\,\mathrm{pkg}}, {\vec{q}}_{s,c,2}^{\,\mathrm{pkg}},
{\vec{q}}_{s,c,3}^{\,\mathrm{pkg}}, {\vec{q}}_{c}^{\,\mathrm{sig}}, 
{\vec{q}}_{c}^{\,\mathrm{ch}})
\end{split}
\end{equation}
as
\begin{widetext}
\begin{equation}
\begin{split}
& P_{s,c}(\dm_k,\vec{x}_{k}\,;\vec{\eta}) = 
 f_{s,c}^{\,\mathrm{comb}} {\cal F}^{\,\mathrm{comb}}(\dm\,;\vec{p}_{s,c}^{\,\mathrm{comb}}) \,
{\cal G}^{\,\mathrm{comb}}(\vec{x}_{k}\,;\vec{q}_{s,c}^{\,\mathrm{comb}}) 
 + (1 - f_{s,c}^{\,\mathrm{comb}})  {\cal F}^{\,\mathrm{peak}}(\delta
m\,;\vec{p}_{c}^{\,\mathrm{peak}}) \\
& \qquad \qquad \times
\left\{ \sum_{j=1}^3\, f_{s,c,j}^{\,\mathrm{pkg}} 
{\cal G}_{j}^{\,\mathrm{pkg}}(\vec{x}_{k}\,;\vec{q}_{s,c,j}^{\,\mathrm{pkg}}) +
\left(1 - \sum_{j=1}^3\,
f_{s,c,j}^{\,\mathrm{pkg}}\right) 
\left[ (1-\fBp)
{\cal G}^\mathrm{sig}(\vec{x}_{k}\,;\vec{q}_{c}^\mathrm{\, sig})
+ \fBp
{\cal G}^\mathrm{ch}(\vec{x}_{k}\,;\vec{q}_{c}^\mathrm{\, ch})
\right]
\right\} \,,
\end{split}
\label{eq:master}
\end{equation}
\end{widetext}
where \dm is the mass difference \massdiff defined earlier.
The symbol ``comb'' in the first term signifies combinatorial-\dst
background.
In the second term,
the symbol ``pkg'' denotes peaking background and 
$j$ indexes the three sources of peaking background (continuum,
fake-lepton and uncorrelated-lepton).
In the last term,
the parameter \fBp describes the charged-$B$ fraction in the sample
after all other types of background are subtracted, and
``sig'' and ``ch'' label functions and parameters for the signal and
charged-$B$ background, respectively. The charged-$B$ fraction is
assumed to be identical for all categories.
The index $g$ is $+1$ ($-1$) for unmixed (mixed) events.
By allowing different effective mistag
rates for apparently mixed or unmixed events
in the background functions ${\cal G}^{\,\mathrm{comb}}$ and ${\cal
G}^{\,\mathrm{pkg}}$, we accomodate the different levels of
backgrounds observed in mixed and unmixed samples. 
Functions labeled with ${\cal F}$ describe the probability of
observing a particular value of $\dm$ while functions labeled with
${\cal G}$ give probabilities for values of $\Dt$ and $\sigmaDt$
in category $g$.
Parameters labeled with $f$
describe the relative contributions of different types of
events. Parameters labeled with $\vec{p}$ describe the shape of a
$\dm$ distribution, and those labeled with $\vec{q}$ describe a
(\Dt, \sigmaDt) shape. The parameters labeled with $\vec{p}$ and $f$
have been determined by a set of fits to \massdiff distributions
described in Sec.~\ref{sec:evtsel}, and are kept fixed in the fit to
(\Dt, \sigmaDt).

Note that we make explicit assumptions that the $\dm$ peak
shape, parameterized by $\vec{p}_{c}^{\,\mathrm{peak}}$, and the
signal and charged-$B$ background
(\Dt, \sigmaDt) shapes, parameterized by
$\vec{q}_{c}^{\,\mathrm{sig}}$ and $\vec{q}_{c}^{\,\mathrm{ch}}$,
depend only on the subsample index $c$ and not on the control sample
index $s$.
The first of these assumptions is
supported by data, and simplifies the analysis of peaking background
contributions. The second assumption reflects our expectation that the
\Dt distribution of signal and charged-$B$ background events does not
depend on whether they are 
selected in the signal sample or appear as a background in a control sample.

The ultimate aim of the fit is to obtain the \Bz lifetime and mixing
frequency, which by construction are common to all sets of signal
parameters $\vec{q}_{c}^{\,\mathrm{sig}}$. Most of the statistical power for
determining these parameters comes from the signal sample, although
the fake and uncorrelated background control samples also contribute
due to their signal content (see Table~\ref{tab:yields}).

We bootstrap the full fit with a sequence of initial fits using reduced
likelihood functions restricted
to a partial set of samples, to determine the
appropriate parameterization of the signal resolution function and the
background \Dt models, and to determine starting values for each
parameter in the full fit. 

\begin{enumerate}
\item
We first find a model that describes the \Dt distribution for each type of 
event: signal, combinatorial-\dst background, and the 
three types of backgrounds that 
peak in the \massdiff distribution.
To establish a model, we use Monte Carlo samples that have been selected
to correspond to only one type of signal or background event
based on the true Monte Carlo information.
These samples are used to determine the \Dt model and 
the categories of events ({\it e.g.,}
tagging category, fake or real lepton) that 
can share each of the parameters in  the model. 
Any subset of parameters can be shared among any subset of the 360
subsamples. We choose parameterizations and sharing of parameters that
minimizes the number of different  parameters while still providing an
adequate description of the \Dt distributions.
\item
We then find the starting values for  the background 
parameters by fitting to each of the background-enhanced control samples
in data, using the model (and sharing of parameters) determined 
in the previous step.
Since these background control samples are not pure, we start with the purest
control sample (combinatorial-\dst background events from the \massdiff
sideband) and move on to less pure control samples, always using the models
established from earlier steps to describe the \Dt distribution of 
the contamination from other backgrounds.
\end{enumerate}

The result of the above two  steps is a \Dt model for each type of event and
a set of starting values for all parameters in the fit.
When we do the final fit, we fit all signal and control samples
simultaneously (approximately 68,000 events), leaving all parameters
in the ${\cal G}$ functions
free in the fit, except for a few parameters that either are
highly correlated with other parameters or reach their physical
limits. The total number of parameters that are free in the fit is 72.
The physics parameters \tauBz and \Dm were kept hidden 
until all analysis details and the systematic errors 
were finalized, to minimize experimenter's bias.
However, statistical errors on the parameters and changes 
in the physics parameters due to changes in the analysis were not hidden.

%%%%%%%%%%%%%%%%%%%%%%%%%%%%%%%%%%%%%%%%%%%%%

\section{Signal \Dt model}

\label{sec:sigmodel}

For signal events in 
a given tagging category $c$, the probability density function (PDF)
for \Dt
consists of a model of the intrinsic time dependence 
convolved with a \Dt resolution function:
\begin{equation}
\begin{split}
 & {\cal G}^{\mathrm{sig}}(\Dt,\sigmaDt,g\,;\vec{q}_c^{\,\mathrm{sig}}) = \\
 & \quad \left\{ \frac{1}{4\tauBz}
e^{-|\Dttrue|/\tauBz}\left[ 1+ g (1-2 \omega_c)\cos(\Dm\Dttrue)\right]\right\} \\
 & \quad \quad  \otimes
{\cal R}(\dDt,\sigmaDt; \vec{q}_c^{\,\mathrm{sig}}) \, ,
\end{split}
\label{eq:sigmodel}
\end{equation}
where ${\cal R}$ is a resolution function, which can be different for
different event categories, 
$g$ is +1 ($-1$) for unmixed (mixed) events,
\dDt is the residual $\Dt - \Dttrue$, and $\omega_c$ is the mistag
fraction for category $c$.
To account for an observed correlation between the mistag rate and
\sigmaDt in the kaon category (described in Sec.~\ref{sec:verttag}),
we allow the mistag rate in the kaon category to vary as a linear
function of \sigmaDt:
\begin{equation}
\omega_\mathrm{\kaon}= \alpha_\mathrm{\kaon}  \sigmaDt +
\omega\mathrm{_{\kaon}^{offset}}\, ,
\label{eq:mistag}
\end{equation}
and allow both the slope $\alpha_\mathrm{\kaon}$ and the offset 
$\omega\mathrm{_{\kaon}^{offset}}$ to be free parameters.
In addition, we allow the
mistag fractions for \Bz tags and \Bzbar tags to be different. 
We define $\Delta \omega= \omega_{\Bz}- \omega_{\Bzbar}$ and 
$\omega=(\omega_{\Bz}+ \omega_{\Bzbar})/2$, so that 
\begin{equation}
\omega_{\Bz / \Bzbar} = \omega \pm \frac{1}{2}\Delta \omega \, .
\end{equation}
The model for the intrinsic time dependence has 13 parameters:
$\omega_c$ and $\Delta \omega_c$ for each of the five
tagging categories, $\alpha_\mathrm{\kaon}$, \Dm and \tauBz.

For the \Dt resolution model,
we use the sum of a single Gaussian distribution and the same
Gaussian convolved with a one-sided exponential to describe
the core part of the resolution function, 
plus a single Gaussian distribution to describe the contribution of 
``outliers'' --- events in which the reconstruction error \dDt is not 
described by the calculated uncertainty \sigmaDt:
\begin{equation}
\begin{split}
& {\cal
R}_\mathrm{GExp+G}(\dDt,\sigmaDt;s,\kappa,f,b^\mathrm{out},
s^\mathrm{out},f^\mathrm{out}) \\  
& = f  G(\dDt; 0,s\sigmaDt) \\
& \quad + (1-f-f^\mathrm{out}) 
G(u-\dDt; 0,s\sigmaDt)\otimes E(u;\kappa\sigmaDt) \\ 
& \quad + f^\mathrm{out} G(\dDt; b^\mathrm{out},s^\mathrm{out})\; ,
\end{split}
\label{eq:GExp}
\end{equation}
where $u$ is an integration variable in the convolution $G\otimes E$.
The functions $G$ and $E$ are defined by 
\begin{equation}
G(x;x_0,\sigma) \equiv {1\over\sqrt{2\pi}\sigma}
  \exp\left[-(x-x_0)^2/(2\sigma)^2\right]
\end{equation}
and
\begin{equation}
E(x;a) \equiv 
 \left\{ \begin{array}{ll}
 {1\over a}\exp{(x/a)} & {\rm if}\ x\leq0, \\
 0           & {\rm if}\ x>0.\end{array}
 \right.
\end{equation}
The exponential component is used to accommodate a bias due to
tracks from charm decays on the \btag side.

Since the outlier contribution is not expected to be described by the
calculated  
error on each event,
the last Gaussian term in Eq.~\ref{eq:GExp} does not depend on \sigmaDt.
However, in the terms that describe the core of the resolution function 
(the first two terms on the right-hand side of Eq.~\ref{eq:GExp}), 
the Gaussian width $s$ and the constant $\kappa$ in the exponential
are scaled by
\sigmaDt. 
The scale factor $s$ is introduced to accommodate an overall underestimate
($s>1$) or overestimate ($s<1$) of the errors for all events.
The constant $\kappa$ is introduced to account for residual charm decay products 
included in the \btag vertex; $\kappa$ is scaled by \sigmaDt to account for
a correlation observed in Monte Carlo simulation  between the mean of the \dDt distribution
and the measurement error \sigmaDt.  

The correlation between \dDt and \sigmaDt 
is due to the fact that, in $B$ decays, the vertex 
error ellipse for the $D$ decay products is oriented with its major axis along the $D$
flight direction, leading to a correlation between the $D$ flight direction and the
calculated uncertainty on the vertex position in $z$ for the \btag candidate.
In addition, the flight length of the $D$ in the $z$ direction is correlated with its
flight direction.
Therefore, the bias in the measured \btag position due to including $D$ decay
products is correlated with the $D$ flight direction.
Taking into account these two correlations, we conclude that $D$ mesons that have a
flight direction perpendicular to the $z$ axis in the laboratory frame will have the
best $z$ resolution and will introduce the least bias in a measurement of the $z$
position of the \btag vertex, while $D$ mesons that travel forward in the laboratory
will have poorer $z$ resolution and will introduce a larger bias in the measurement of
the \btag vertex.

The mean and RMS spread of \Dt residual distributions in Monte Carlo simulation
vary significantly among tagging categories.
We find that we can account for these differences by allowing the fraction of 
core Gaussian, $f$, to be different for each tagging category.
In addition, we find that the correlations among the three parameters 
describing the outlier Gaussian ($b^\mathrm{out}$, $s^\mathrm{out}$, $f^\mathrm{out}$)
are large and that the outlier parameters are highly correlated with other
resolution parameters. 
Therefore, we fix the outlier bias $b^\mathrm{out}$ and width 
$s^\mathrm{out}$, and vary them over a wide range to evaluate 
the systematic uncertainty on the physics parameters due to fixing these parameters
(see Sec.~\ref{sec:systematics}).
The signal resolution model then has eight free parameters: $s$, $\kappa$,
$f^\mathrm{out}$, 
and five fractions $f_c$ (one for each tagging category $c$).

As a cross-check, we use a resolution function that is the sum of
a narrow and a wide Gaussian distribution, and a third Gaussian to describe 
outliers:
\begin{equation}
\begin{split}
& {\cal R}_\mathrm{G+G+G}(\delta\Dt,\sigmaDt;b,s,f,b^w,s^w,b^\mathrm{out},s^\mathrm{out},f^\mathrm{out}) \\
& =  f  G(\dDt; b\sigmaDt,s\sigmaDt) \\
& \quad + (1-f-f^\mathrm{out})  G(\dDt; b^w\sigmaDt,s^w\sigmaDt) \\
& \quad + f^\mathrm{out}  G(\dDt; b^\mathrm{out},s^\mathrm{out})\; .
\end{split}
\end{equation}
This resolution function has two more parameters than ${\cal
R}_\mathrm{GExp+G}$.
It accommodates a bias due to tracks from charm decays on the
\btag side by allowing the means of the Gaussian distributions to be nonzero.

%%%%%%%%%%%%%%%%%%%%%%%%%%%%%%%%%%%%%%%%%%%%%

\subsection{Vertex-tagging correlations}

\label{sec:verttag}

A correlation $d\omega_c/d\sigma_{\Delta t}\approx 0.12$~ps$^{-1}$ 
is observed between the mistag rate and the 
\Dt resolution for \kaon tags.
This effect is modeled in the resolution function for signal as a linear dependence
of the mistag rate on \sigmaDt, as shown in Eq.~\ref{eq:mistag}.
In this section, we describe the source of this correlation.

We find that both the mistag rate for \kaon tags and the calculated
error on \Dt depend inversely on $\sqrt{\Sigma p_t^2}$, where $p_t$ is
the transverse momentum with respect to the $z$ axis of tracks from
the \btag decay. Correcting for this dependence of the mistag rate
removes most of the correlation between the mistag rate and \sigmaDt.
The mistag rate dependence originates from the kinematics of the
physics sources for wrong-charge kaons. The three major sources of
mistagged events in the \kaon category are wrong-sign \dz mesons from
$B$ decays to double charm ($b\rightarrow c \overline c s$),
wrong-sign kaons from $D^+$ decays, and 
kaons produced directly in $B$ decays. All these sources produce a
spectrum of tracks that have smaller $\sqrt{\Sigma p_t^2}$ than $B$
decays that produce a correct tag. The \sigmaDt dependence originates
from the $1/p^2_t$ dependence of \sigmaz for the individual
contributing tracks.

%%%%%%%%%%%%%%%%%%%%%%%%%%%%%%%%%%%%%%%%%%%%%
%\vspace{1cm}
\section{\Dt models for backgrounds}

\label{sec:bkgndmodel}

Although the true \Dt and resolution on \Dt are not well-defined  
for background events, we still
describe the total \Dt model as a ``physics model''
convolved with a ``resolution function''.

The background \Dt physics models we use in this analysis are
all a linear combination of one or more of the following terms,
corresponding to prompt, exponential 
decay, and oscillatory 
distributions:
\begin{widetext}
\begin{eqnarray}
{\cal G}^\mathrm{prmt}_\mathrm{phys}(\Dttrue,g) & = &
{1\over 2}\delta(\Dttrue) \left[ 1 + g  (1- 
\omega^\mathrm{prmt})\right] \, ,\\
{\cal G}^\mathrm{life}_\mathrm{phys}(\Dttrue,g) & = & 
{1\over 4\tau^\mathrm{bg}}
\exp({-|\Dttrue|/\tau^\mathrm{bg}}) \left[ 1 +
g (1-\omega^\mathrm{life})\right] \, ,\\ 
{\cal G}^\mathrm{osc}_\mathrm{phys}(\Dttrue,g) & = & 
{1\over 4\tau^\mathrm{bg}}
\exp({-|\Dttrue|/\tau^\mathrm{bg}}) \left[ 1 +
g (1-\omega^\mathrm{osc})\cos(\Delta m^\mathrm{bg} \Dttrue) \right] \, ,
\end{eqnarray}
\end{widetext}
where $\delta(\Dt)$ is a $\delta$-function, 
$g=+1$ for unmixed and $-1$ for mixed events, 
and $\tau^\mathrm{bg}$ and $\Delta m^\mathrm{bg}$ are the
effective lifetime and mixing frequency for the particular background.

For backgrounds, we use a resolution function that is the sum of
a narrow and a wide Gaussian distribution:
\begin{equation}
\begin{split}
& {\cal R}_\mathrm{G+G}(\dDt,\sigmaDt;b,s,f,b^w,s^w) \\ 
& = f G(\dDt; b\sigmaDt,s\sigmaDt) +  
(1-f) G(\dDt; b^w\sigmaDt,s^w\sigmaDt)\,.
\end{split}
\end{equation}
 
\subsection{Combinatorial-\dst background}
\label{sec:combbkgnd}

Events in which the \dstm candidate corresponds to a random combination
of tracks (called combinatorial-\dst background) constitute the largest
background in the signal sample.
We use two sets of events to determine the appropriate parameterization of 
the \Dt model for combinatorial-\dst background:
events in data that are in the upper \massdiff sideband (above the
peak due to real \dstm decays); and 
events in Monte Carlo simulation that are identified as combinatorial-\dst
background, 
based on the true information for the event, in both the \massdiff
sideband and peak region. 
The data and Monte Carlo \Dt distributions are described well by a
prompt plus oscillatory term convolved with a double-Gaussian
resolution function: 
\begin{equation}
\begin{split}
{\cal G}^\mathrm{comb} = & \Bigl[ f^\mathrm{osc}  
{\cal G}\mathrm{_{phys}^{osc}}
(\Dttrue,g;\tau^\mathrm{comb},\Delta
m^\mathrm{comb},\omega^\mathrm{osc}) \\
 & \;\; + 
(1- f^\mathrm{osc}) 
{\cal G}\mathrm{_{phys}^{prmt}}
(\Dttrue,g;\omega^\mathrm{prmt}) \Bigr] \otimes 
{\cal R}_\mathrm{G+G} \;.
\end{split}
\label{eq:combdt}
\end{equation}

Approximately 60\% of combinatorial-\dst 
background events are from $\Bz\Bzbar$
events according to Monte Carlo simulation. Although the \dstm is not
correctly reconstructed, the identified lepton is very likely to
be a primary lepton. The tagging algorithm can still identify the
flavor of \btag with a reasonable mistag probability, especially for
the \lepton category, and for the \kaon category if the tracks swapped
between the \dstl candidate and \btag are pions. Therefore, the
combinatorial-\dst background also exhibits oscillatory behavior.

The parameters $\omega^\mathrm{prmt}$, $\Delta m^\mathrm{comb}$,
$\tau^\mathrm{comb}$,  
$f$, $b^w$, and $s^w$ are shared among all subsamples.
The parameters $\omega^\mathrm{osc}$, $f^\mathrm{osc}$, $b$, and $s$ 
are allowed
to be different depending on criteria such as tagging category, 
whether the data were recorded on- or off-resonance,
whether the candidate lepton passes real- or fake-lepton criteria,
and
whether the event passes the criteria for same-side or opposite-side
\dstm and $\ell$.
The total number of free parameters in the combinatorial-\dst 
background \Dt model is 24. 

The relative fraction of $\Bz\Bzbar$ and $B^+B^-$ events in the 
combinatorial-\dst background depends slightly on \massdiff. 
However, no significant dependence of the parameters
of the \Dt model on \massdiff is observed in data or Monte Carlo simulation.
The sample of events in the \massdiff sideband is used to determine
the starting 
values for the parameters in the final full fit to all data samples.

To reduce the total number of free parameters in the fit, parameters 
that describe 
the shape of the wide Gaussian (bias and width) 
are shared between combinatorial-\dst 
background and the three types of peaking background: 
continuum, fake-lepton, and uncorrelated-lepton.
The wide Gaussian fraction is allowed to be different for each type of
background. 

\subsection{Continuum peaking background}
\label{sec:contbkgnd}

All $c\overline c$ events that have a correctly
reconstructed \dstm are defined as continuum peaking background,
independent of whether the associated lepton 
candidate is a real lepton or a fake lepton.
The $c\overline c$ Monte Carlo sample and off-resonance data are 
used to identify the appropriate \Dt model and sharing of parameters
among subsamples.
The combinatorial-\dst background \Dt model and parameters described in the 
previous section are used to model the combinatorial-\dst background
contribution in the off-resonance \Dt distribution in data.

The decay vertex of a real \dstm from continuum $c\overline c$
production always coincides with the primary vertex. If the lepton
candidate also originates from the primary vertex, we can use a
prompt physics model convolved with a resolution function that
can accommodate a bias due to tracks from charm decays other than the
\dstm candidate. If the lepton candidate is from a charm decay,
the measured vertices of the \dstl candidate and the remaining tracks
are both likely to be between the primary vertex and the charm vertex;
hence the measured \Dz is likely to be very small.
Both types of events can be modeled with a prompt model
convolved with a double-Gaussian resolution function:
\begin{equation}
 {\cal G}^\mathrm{cont} 
={\cal G}^\mathrm{prmt}_\mathrm{phys}(\Dttrue,g;\omega^\mathrm{prmt})
\otimes
{\cal R}_\mathrm{G+G}\; .
\end{equation}
Dependence on the flavor tagging information is included to accomodate
any differences in the amount of background events classified as mixed
and unmixed. 

By fitting to the data and Monte Carlo control samples
with different sharing of parameters across subsets of the data, 
we find that the apparent ``mistag fraction'' for events  in the \kaon
category is significantly different from the 
mistag fraction for other tagging categories.
We also find that the core Gaussian bias 
is significantly different for opposite-side and same-side events.
We introduce separate parameters to accommodate these effects.

The total number of parameters used to describe the \Dt distribution of 
continuum peaking background is six.
The off-resonance control samples in data are  
used to determine starting values for the final full fit to all data samples.

\subsection{Fake-lepton peaking background}
\label{sec:fakebkgnd}

To determine the \Dt model and sharing of parameters for the fake-lepton 
peaking backgrounds, we use $\Bz\Bzbar$ and $B^+B^-$ Monte Carlo
events in which 
the \dstm is correctly reconstructed 
but the lepton candidate is misidentified.
In addition, we use the fake-lepton control sample in data. 
The combinatorial-\dst and continuum peaking background \Dt models and
parameters described in the  previous two sections are used to model
their contribution to the fake-lepton \Dt distribution in data.
For this study, the contribution of signal is described by the signal
parameters found for signal events in the Monte Carlo simulation.

Since the fake-lepton peaking background is due to $B$ decays in which
the fake lepton and the \dstm candidate can originate from the same $B$ or 
different $B$ mesons, and the charge of the fake lepton can carry
correct flavor information of the reconstructed $B$ candidate,
we include both prompt and oscillatory terms in the \Dt model:
\begin{equation}
\begin{split}
{\cal G}^\mathrm{fake} = & \Bigl[ f^\mathrm{osc}  
{\cal G}\mathrm{_{phys}^{osc}}
(\Dttrue,g;\tau^\mathrm{fake},\Delta
m^\mathrm{fake},\omega^\mathrm{osc}) \\
 & \;\, + 
(1- f^\mathrm{osc}) 
{\cal G}\mathrm{_{phys}^{prmt}}
(\Dttrue,g;\omega^\mathrm{prmt}) \Bigr] \otimes 
{\cal R}_\mathrm{G+G} \;.
\end{split}
\end{equation}
We find that the apparent mistag rates for both the prompt and mixing terms, and the
bias of the core Gaussian of the resolution function, are different between some tagging
categories. 
The total number of parameters used to describe the fake-lepton background is 14.
The fake-lepton control samples in data are  
used to determine starting values for the final full fit to all data samples.

\subsection{Uncorrelated-lepton peaking background}
\label{sec:uncorrbkgnd}

To determine the \Dt model and sharing of parameters for the uncorrelated-lepton 
peaking backgrounds, we use $\Bz\Bzbar$ and $B^+B^-$ Monte Carlo events in which
the \dstm is correctly reconstructed but the lepton candidate is from the other
$B$ in the event or from a secondary decay of the same $B$.
In addition, we use the same-side control sample in data, 
which is only about 30\%
uncorrelated-lepton background 
in the \massdiff peak region due to significant contributions from 
combinatorial-\dst  background
and signal.
The combinatorial-\dst and peaking background \Dt models and
parameters described 
in the  previous two sections are used to model their
contribution to the same-side \Dt distribution in data.
For this initial fit, the contribution of signal is described by the signal
parameters found for signal events in the Monte Carlo simulation.

Physics and vertex reconstruction considerations suggest several features of the
\Dt distribution for the uncorrelated-lepton sample.  
First, we expect the reconstructed \Dt to be systematically
smaller than the true \Dt value since using a lepton and a \dstm from 
different $B$ decays  
will generally reduce the separation between the reconstructed
\bdstl and \btag vertices. 
We also expect that events with small true \Dt will have a higher
probability of being misreconstructed as an uncorrelated lepton
candidate because it is more likely that the fit of the \dstm and $\ell$ to
a common vertex will converge for these events.
Finally, we expect truly mixed events to have a higher fraction of uncorrelated-lepton
events because in mixed events the charge of the \dst is opposite that of
primary leptons on the tagging side.
These expectations are confirmed in the Monte Carlo simulation.

We do not expect the uncorrelated-lepton background to exhibit any mixing behavior
and none is observed in the data or Monte Carlo control samples.
We describe the \Dt distribution with the sum of a lifetime term and a prompt term, 
convolved with a double-Gaussian resolution function:
\begin{equation}
\begin{split}
{\cal G}^\mathrm{uncor} = & \Big[ f^\mathrm{life}  
{\cal G}\mathrm{_{phys}^{life}}
(\Dttrue,g;\tau^\mathrm{uncor},
\omega^\mathrm{life}) \\
 & \;\, + 
(1- f^\mathrm{life}) 
{\cal G}\mathrm{_{phys}^{prmt}}
(\Dttrue,g;\omega^\mathrm{prmt}) 
\Big] \otimes   
{\cal R}_\mathrm{G+G} \;.
\end{split}
\label{eq:uncodt}
\end{equation}
The effective mistag rates $\omega^\mathrm{prmt}$ and $\omega^\mathrm{life}$
accommodate different fractions
of uncorrelated-lepton backgrounds in events classified as mixed and unmixed.
We find that the apparent mistag
rate for the lifetime term is different between some tagging categories. 
All other parameters are consistent among the different subsamples. 
The total number of parameters used to describe the uncorrelated-lepton background is six.
The uncorrelated-lepton control samples in data are  
used to determine starting values for the final full fit to all data samples.

\subsection{Charged $B$ peaking background}
\label{sec:chBbkgnd}

The charged-$B$ peaking background is due to decays of the type 
$B^\pm\rightarrow \dst\ell\nu_\ell X$.
Since charged $B$'s do not exhibit mixing behavior,
we use the \Dt and tagging information
to discriminate charged-$B$ peaking background events
from neutral-$B$ signal events, in the simultaneous fit to all samples. 
We use the same resolution model and parameters as for the  
neutral-$B$ signal since the \Dt resolution is dominated by the
$z_\mathrm{tag}$ resolution and
the $B$ decay dynamics are very similar.
The charged $B$ background contribution is described by
\begin{equation}
\begin{split}
{\cal G}^{\mathrm{ch}} = 
& \frac{1}{4\tauBp} e^{-|\Dttrue|/\tauBp}\left[1+ g (1-2 \omega_{B^+}^c)
\right] \\
& \;\; \otimes {\cal R}(\dDt,\sigmaDt; \vec{q}_c) \, ,
\end{split}
\end{equation}
where $\omega_{B^+}^c$ is the mistag fraction for
charged $B$ mesons for tagging category  $c$.

Given that the ratio of the charged $B$ to neutral $B$ lifetime is close to
1 and the fraction of charged $B$ mesons in the peaking sample is small, we
do not have sufficient sensitivity to distinguish the lifetimes in the fit. 
We parameterize the physics model for the $B^+$ in terms of the 
lifetime ratio 
$\tauBp/\tauBz$, and fix this ratio to the Review of Particle Properties 2002 world average of 1.083\cite{ref:PDG2002}.
We vary the ratio by the error on the world average ($\pm0.017$) to 
estimate the corresponding systematic
uncertainties on \tauBz and \Dm (see Sec.~\ref{sec:systematics}).

In each tagging category, 
the fit is sensitive to only two parameters 
among \wBp, the neutral $B$ mistag fraction (\wBz) and the charged $B$
fraction (\fBp). 
Therefore we fix the ratio  of mistag rates, $\wBp/\wBz$,
to the value of the ratio measured with fully reconstructed charged and 
neutral hadronic $B$ decays in data, for each tagging category.

%%%%%%%%%%%%%%%%%%%%%%%%%%%%%%%%%%%%%%%%%%%%%

\section{Fit results}

\label{sec:results}

The total number of free parameters in the final fit is 72: 
21 in the signal model, one for the charged $B$ fraction,
24 in the combinatorial-\dst background model, 
and 26 in peaking background models. 
The fitted signal \Dt model parameters are shown in
Table~\ref{tab:result-signal}. 

\begin{table}[htb]
\caption{Results of full fit to data --- signal model and 
resolution function parameters.
A correction, described in Sec.~\ref{sec:MCtests}, has been
applied to \tauBz and \Dm. The uncertainties are statistical only.}
\begin{center}
\begin{tabular*}{0.48\textwidth}{@{\extracolsep{\fill}}lclc}
\hline\hline
 parameter & value & parameter & value \\
\hline
\Dm (ps$^{-1}$) & $ 0.492\pm 0.018$ & $\Delta \omega_\mathrm{\nttwo}$ & $-0.112\pm 0.028$ \\
\tauBz (ps) & $ 1.523\,^{+0.024}_{-0.023}$ & $\Delta
\omega_\mathrm{\ntthree}$ & $-0.023\pm 0.019$ \\
\fBp & $ 0.082\pm 0.029$ & $s$ & $ 1.201\pm 0.063$ \\ 
$\omega_\mathrm{\lepton}$ & $ 0.071\pm 0.015$ & $\kappa$ & $ 0.86\pm 0.17$ \\ 
 $\omega\mathrm{_{\kaon}^{offset}}$ & $ 0.002\pm 0.024$ & $f\mathrm{_{\lepton}}$ & $ 0.72\pm 0.10$ \\ 
 $\alpha_\mathrm{\kaon}$ (ps$^{-1}$) & $ 0.229\pm 0.036$ & $f_\mathrm{kaon}$ & $ 0.609\pm 0.088$ \\ 
 $\omega_\mathrm{\ntone}$ & $ 0.212\pm 0.020$ & $f\mathrm{_{\ntone}}$ & $ 0.69\pm 0.13$ \\ 
 $\omega_\mathrm{\nttwo}$ & $ 0.384\pm 0.018$ & $f\mathrm{_{\nttwo}}$ & $ 0.70\pm 0.10$ \\ 
 $\omega_\mathrm{\ntthree}$ & $ 0.456\pm 0.012$ & $f\mathrm{_{\ntthree}}$ & $ 0.723\pm 0.078$ \\ 
 $\Delta \omega_\mathrm{\lepton}$ & $-0.001\pm 0.022$ & $f\mathrm{^{out}}$ & $ 0.0027\pm 0.0017$ \\ 
 $\Delta \omega_\mathrm{kaon}$ & $-0.024\pm 0.015$ &
$b\mathrm{^{out}}$ (ps) & $-5.000$ (fixed) \\ 
 $\Delta \omega_\mathrm{\ntone}$ & $-0.098\pm 0.032$ & $s\mathrm{^{out}}$ (ps) & $ 6.000$ (fixed)\\ 
\hline\hline
\end{tabular*}
\end{center}
\label{tab:result-signal}
\end{table}

The statistical correlation coefficient 
between \tauBz and \Dm 
is $\rho(\Dm,\tauBz)= -0.22$. 
The global correlation coefficients (the largest correlation between
a variable and every possible linear combination of other variables) 
 for \tauBz and \Dm, and
some of the correlation
coefficients between
\tauBz or \Dm and other parameters, are 
shown in Table~\ref{tab:result-cor}.

\begin{table}[htb]
\caption{Global correlation coefficients for 
\Dm and \tauBz from the full fit to data and other
correlation coefficients for pairs of key parameters
in the fit.}
\begin{center}
\begin{tabular*}{0.48\textwidth}{@{\extracolsep{\fill}}lr}
\hline\hline
\Dm global correlation &  0.74 \\
\tauBz global correlation &  0.69 \\
\\
$\rho(\Dm,\tauBz)$ & $-0.22$ \\
$\rho(\Dm,\fBp)$ & 0.58 \\
$\rho(\tauBz,s_\mathrm{sig})$ & $-0.49$ \\
$\rho(\tauBz,f\mathrm{_{sig}^{out}})$ & $-0.26$ \\
\hline\hline
\end{tabular*}
\end{center}
\label{tab:result-cor}
\end{table}

\begin{figure*}[htb]
\begin{center}
\includegraphics[width=2.5in]{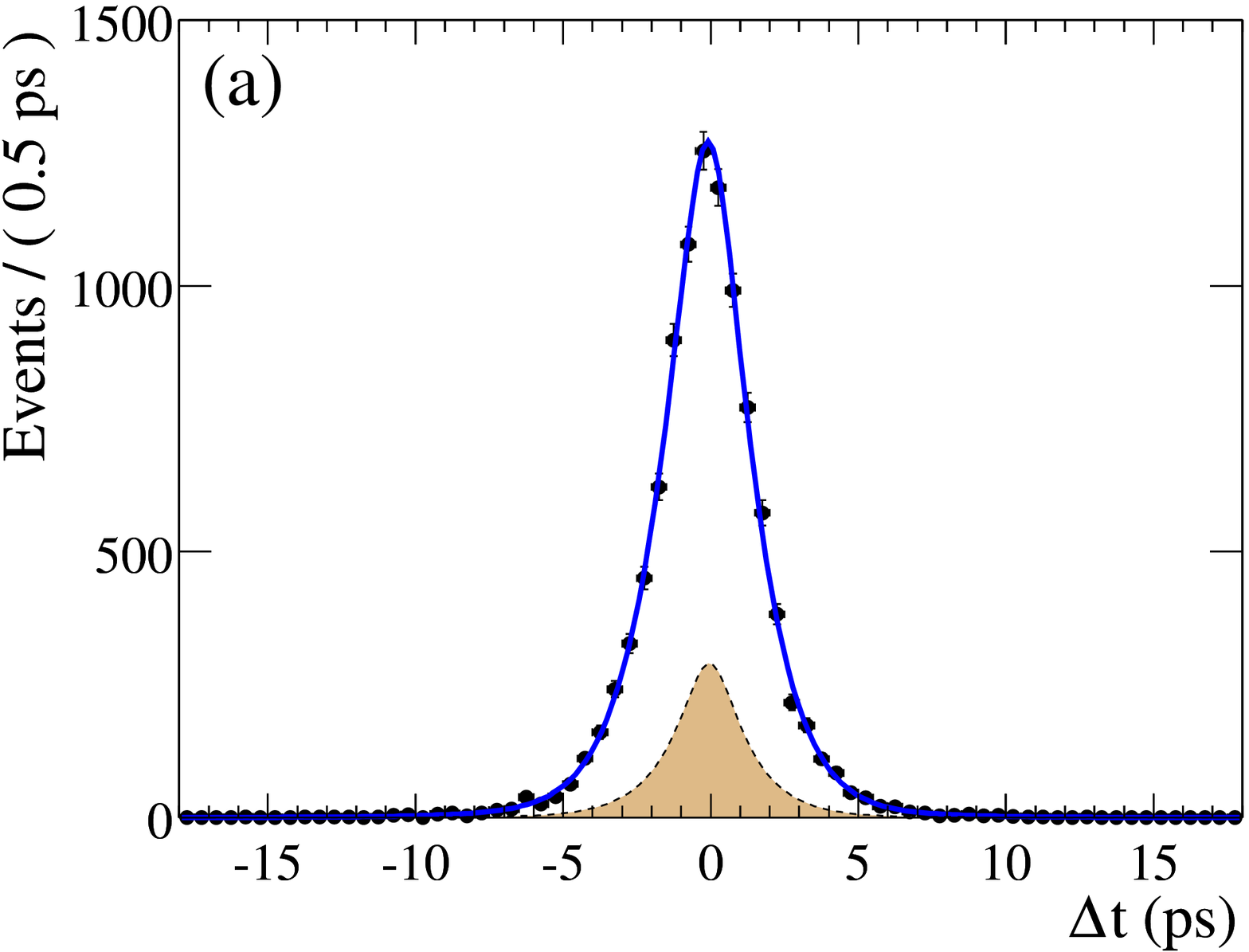}
\includegraphics[width=2.5in]{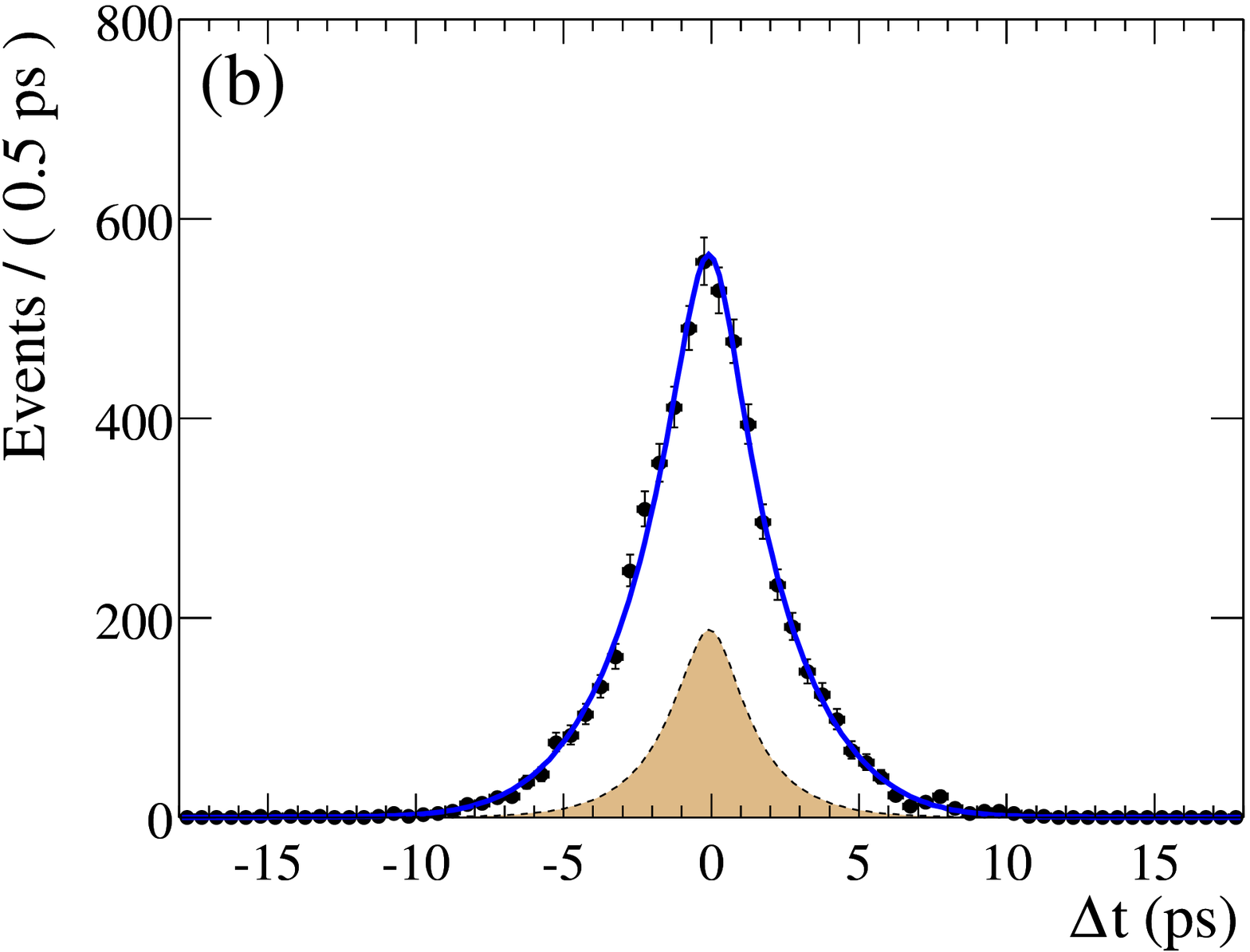} \\
\includegraphics[width=2.5in]{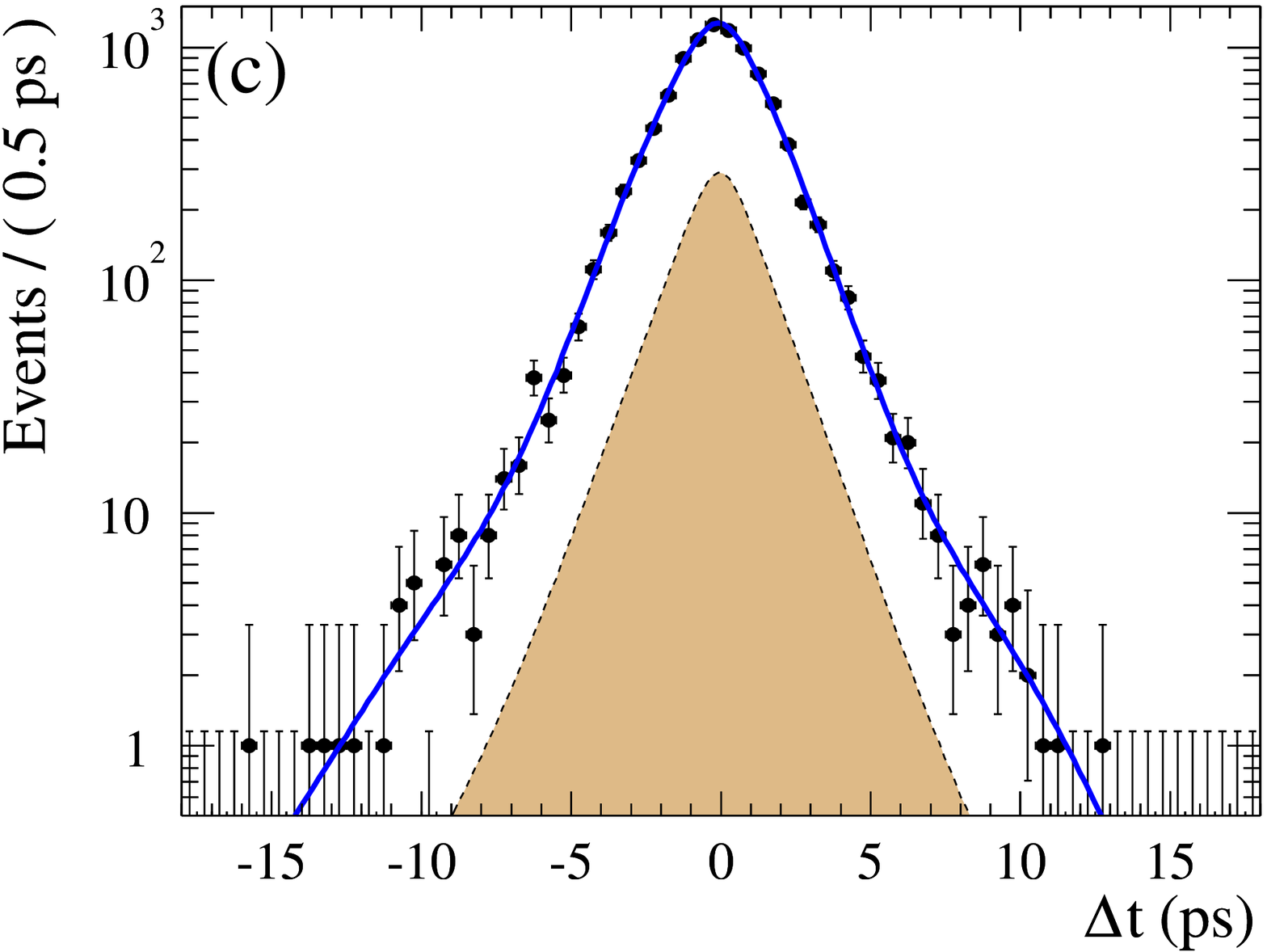}
\includegraphics[width=2.5in]{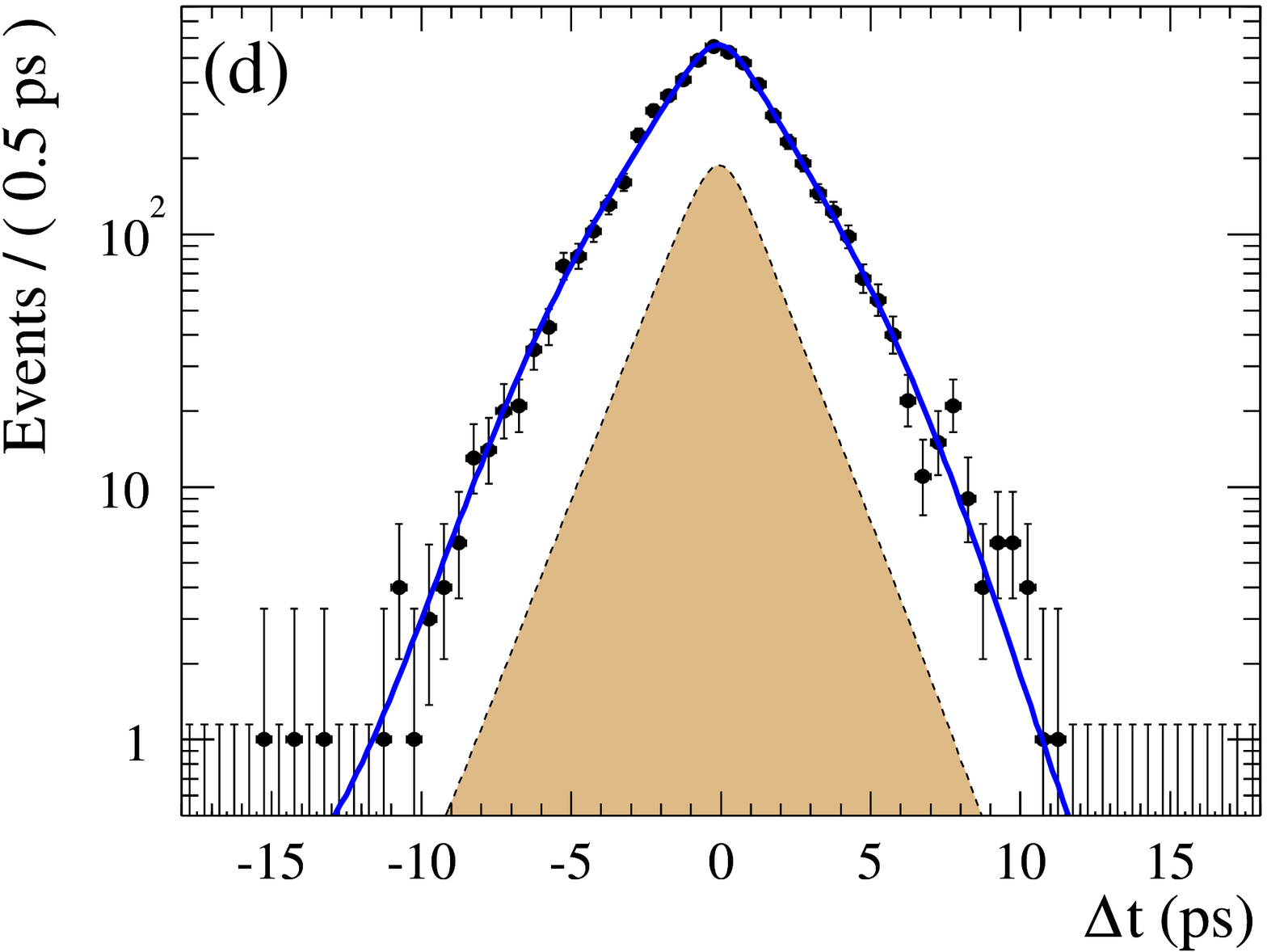}
\end{center}
\caption{The \Dt distribution on linear (a, b) and logarithmic (c, d)
 scale for (a, c) unmixed and (b, d) mixed events in an 
80\%-pure signal sample and the projection of the fit results.
Each event in this sample has a probability of being a signal 
higher than a threshold chosen so that the sample 
is 80\% pure in signal events.
The shaded area shows the background contribution to
the distributions.} 
\label{fig:result-dt}
\end{figure*}

\begin{figure}[htb]
\begin{center}
\includegraphics[width=3in]{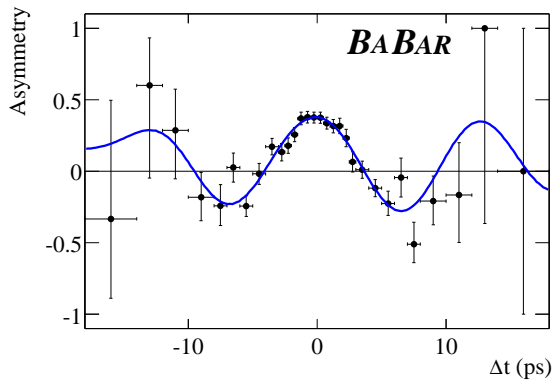}
\end{center}
\caption{The asymmetry plot for mixed and unmixed events in
an 80\%-pure signal sample
and the projection of the fit results. Errors on the data points
are computed by considering the binomial probabilities for 
observing different numbers of mixed and unmixed
events while preserving the total number. }
\label{fig:result-asym}
\end{figure}

Figure~\ref{fig:result-dt} 
shows the \Dt distributions for unmixed and mixed events in 
a sample in which the probability of each event being a signal is
higher than a threshold chosen so that the sample 
is 80\% pure in signal events.
The points correspond to data.
The curves correspond to the sum of the
projections of the appropriate relative amounts of signal and 
background \Dt models for this 80\%-pure signal sample.
Figure~\ref{fig:result-asym} shows the time-dependent asymmetry
\begin{equation}
A(\Dt) =   {N_\mathrm{unmixed}(\Dt) - N_\mathrm{mixed}(\Dt)
  \over  N_\mathrm{unmixed}(\Dt) + N_\mathrm{mixed}(\Dt)}.
\end{equation}
The unit amplitude for the cosine dependence of 
$A$ is diluted by the mistag probabilities, the experimental
\Dt resolution, and backgrounds.

\begin{figure*}[htb]
\begin{center}
\includegraphics[width=2.5in]{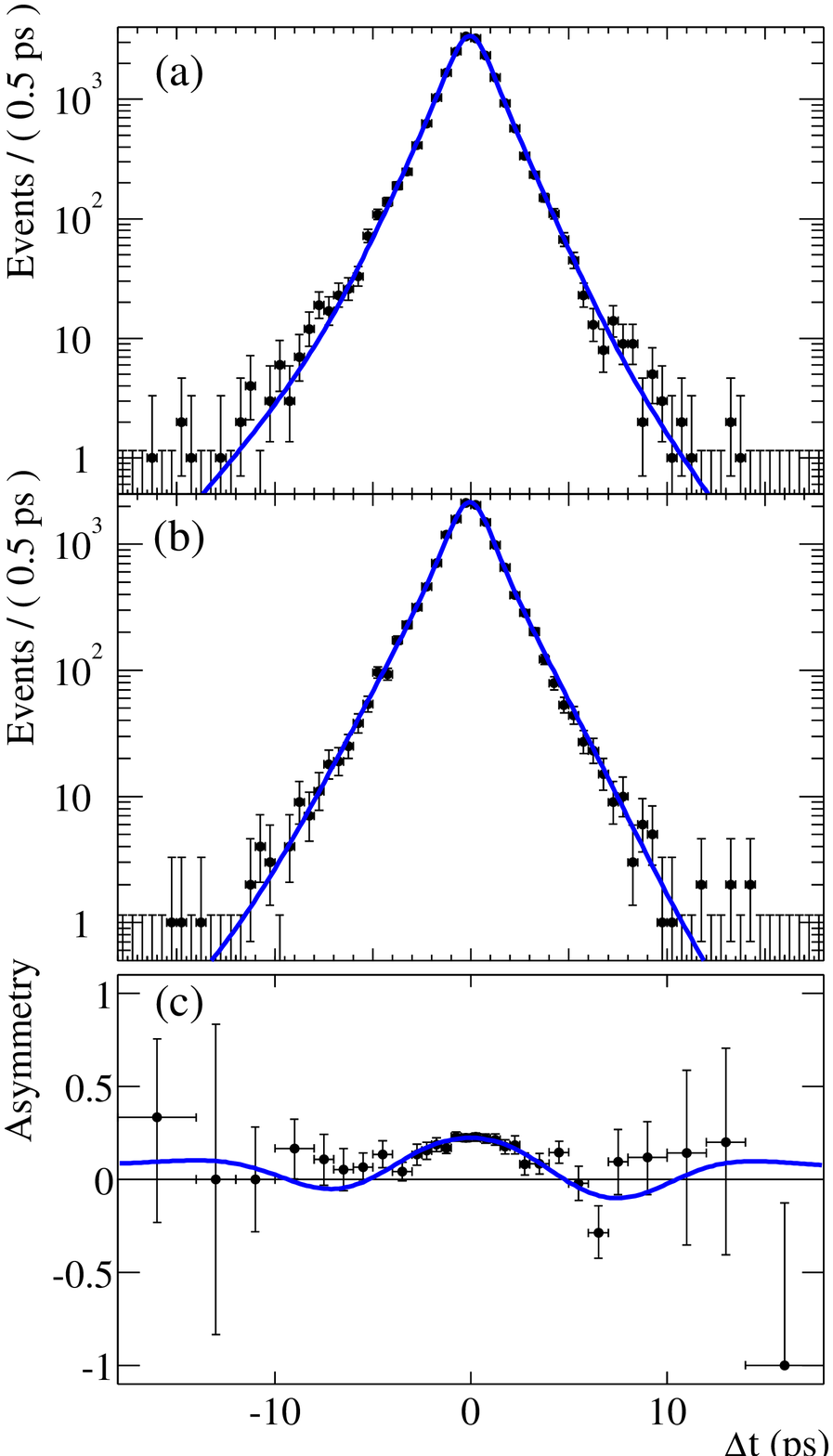}
\includegraphics[width=2.5in]{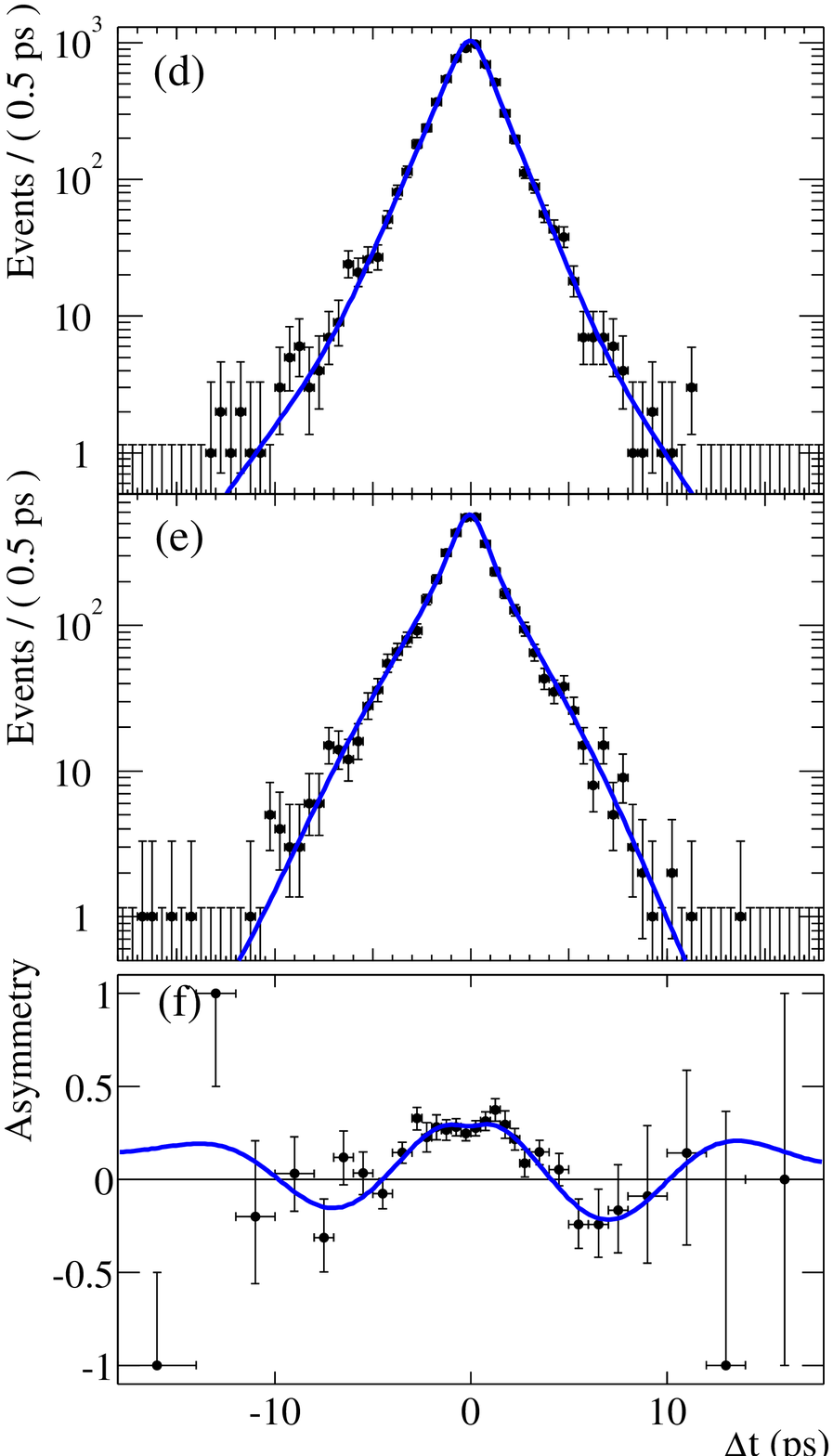}
\end{center}
\caption{The \Dt distributions for (a, d) unmixed and (b, e) mixed
events, and (c, f) the asymmetry plot in a 99.5\%-pure
combinatorial-\dst sample (a, b and c) and in a 60\%-pure \dstm--fake-lepton event
sample (d, e and f). Events are selected based on the background
probabilities, such that the sample contains 99.5\%-pure combinatorial-\dst
events, or 60\%-pure \dstm--fake-lepton background events.
The projection of the fit results is overlayed on top of the data points. 
Errors on the data points in the asymmetry plots
are computed by considering the binomial probabilities for 
observing different numbers of mixed and unmixed
events while preserving the total number. }
\label{fig:result-bgdt}
\end{figure*}

Figure~\ref{fig:result-bgdt} shows the \Dt distributions for unmixed
and mixed events, and the asymmetry $A(\Dt)$ for data samples in which
events are selected based on the background probabilities
such that the sample contains 99.5\%-pure combinatorial background
events (left plots), or 60\%-pure fake-lepton background events
(right plots). The observed oscillatory behaviors are expected as
explained in Sec.~\ref{sec:bkgndmodel}.

\begin{figure}[htb]
\begin{center}
\includegraphics[width=3in]{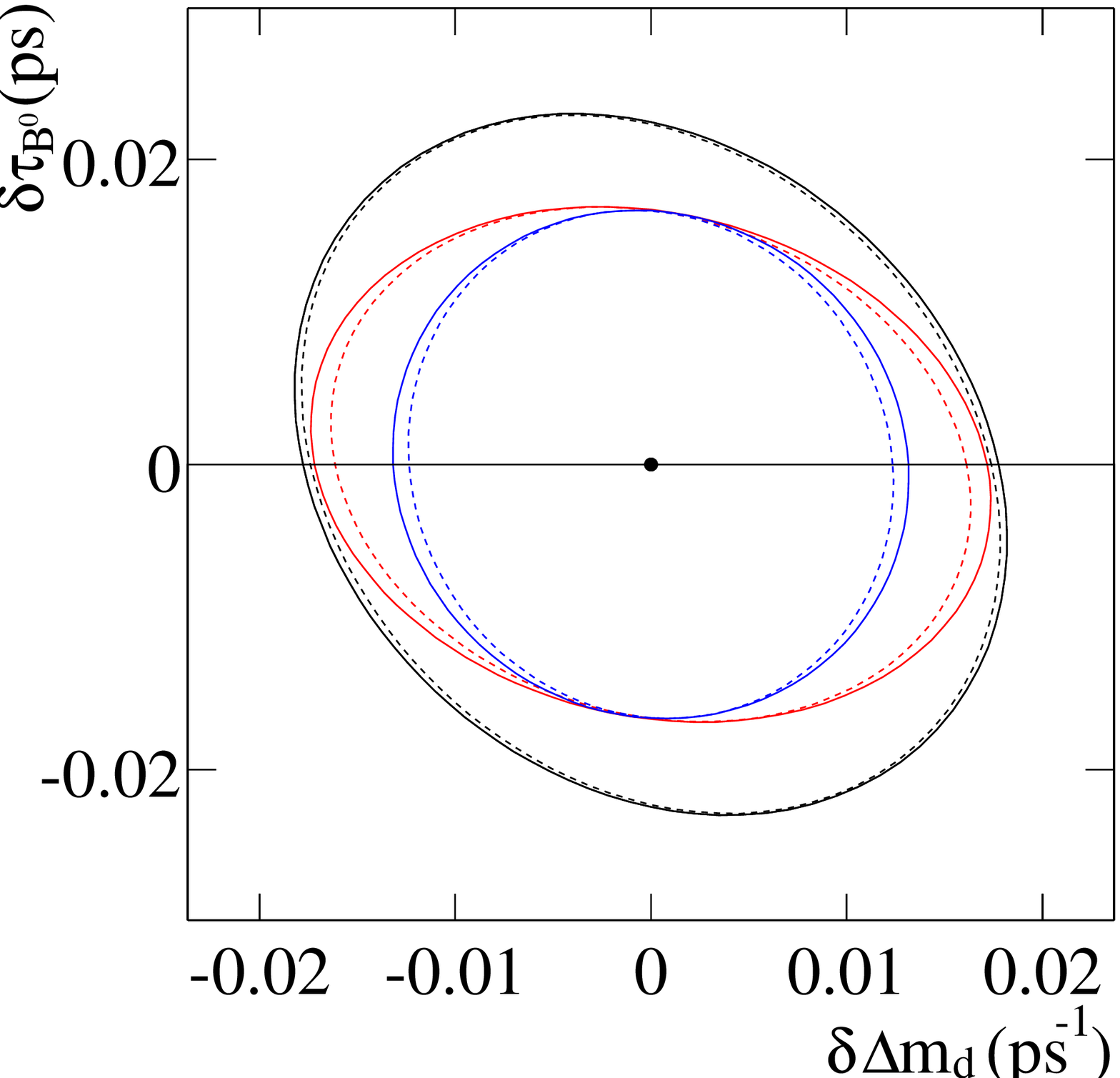}
\end{center}
\caption{Comparison of one-sigma error ellipses in the  \Dm-\tauBz\
plane for fits in which different sets of parameters are free. From
the innermost to the outermost ellipse, the floating parameters are 
(\Dm, \tauBz),
(\Dm, \tauBz, mistag fractions), (\Dm, \tauBz, \fBp), (\Dm, \tauBz,
\fBp, mistag fractions), all
signal \Dt parameters, and the default fit (72 floating parameters).}
\label{fig:contourNest}
\end{figure}

Since many parameters in the model are free, it is interesting to see
how the errors on \tauBz and \Dm, and their correlation, change when 
different parameters are free in the fit, 
or fixed to their best value from the full fit.
We perform a series of fits, fixing all
parameters at the values obtained from the default fit, except (a) \Dm\
and \tauBz, (b) \Dm, \tauBz, and all mistag fractions in the signal model, 
(c) \Dm, \tauBz, and \fBp, (d) \Dm, \tauBz, \fBp, and all mistag fractions
in the signal model, (e) all parameters in the signal \Dt model. 
The one-sigma error ellipses for these fits
and for the default fit are shown in Fig.~\ref{fig:contourNest}.

We can see that the error on \tauBz changes very little until we float
the signal resolution function. Floating the background parameters adds
a very small contribution to the error. 
The contribution from the charged $B$
fraction and mistag fractions to the \tauBz error is negligible. 
On the other hand, the charged $B$ fraction changes the error on \Dm the
most. The contributions from floating the mistag fractions, resolution
functions, and background \Dt models are relatively small.

We also check the statistical errors on data 
by measuring the increase in negative log likelihood 
in data in the two-dimensional (\tauBz, \Dm) space in the vicinity of the 
minimum of the negative log likelihood.  
We find that the positive error on \tauBz is about 6\% larger than 
that determined by the fitting program, whereas the other errors are 
the same as those determined by the fit.
To take this into account, we increase the positive statistical error
on \tauBz by 6\%.

%%%%%%%%%%%%%%%%%%%%%%%%%%%%%%%%%%%%%%%%%%%%%

\section{Validation and cross checks}

\label{sec:validations}

%% Validations

In Sec.~\ref{sec:MCtests}, we describe several tests of the fitting 
procedure that were performed with both fast parameterized Monte Carlo
simulations and full detector simulations.
In Sec.~\ref{sec:datachecks}, we give the results of performing
cross-checks on data, including fitting to different subsamples of the
data and fitting with variations to the standard fit.

\subsection{Tests of fitting procedure with Monte Carlo simulations}
\label{sec:MCtests}

A test of the fitting procedure is performed with fast 
parameterized Monte Carlo simulations, where 87 experiments are 
generated with signal and background control sample sizes and
compositions corresponding to that obtained from the full likelihood fit
to data.   
The mistag rates and \Dt distributions are generated
according to the model used in the likelihood fit.
The full fit is then performed on each of these experiments.
We find no statistically significant bias in
the average values of \tauBz and \Dm for the 87 fits.
The RMS spread in the distribution of results is consistent with
the mean statistical error from the fits and the statistical
error on the results in data, for both \tauBz and \Dm. 
We find that 20\% of the fits result in a value
of the negative log likelihood that is smaller (better) than that found
in data. 

We also fit two types of Monte Carlo samples that include full detector
simulation: pure signal and signal plus background.
To check whether the selection criteria introduce any bias in the 
lifetime or mixing frequency,  we fit the signal physics model to
the true lifetime distribution, using true tagging information, for
a large sample of signal Monte Carlo events that pass all selection
criteria. 
We also fit the measured \Dt distribution, using measured tagging 
information, with the complete signal \Dt model described in 
Sec.~\ref{sec:sigmodel}.
We find no statistically significant bias in the values of \tauBz or 
\Dm extracted in these fits.

The \BzBzbar, $B^+B^-$, and $c\overline c$ Monte Carlo samples that
provide simulated background events along with signal events are much
smaller than the pure signal Monte Carlo samples.
In addition, they are not much larger than the data samples.  
In order to increase the statistical sensitivity to any bias introduced
when the  background samples are added to the fit, we compare the 
values of \tauBz and \Dm from the fit to signal plus background
events, and pure signal events from the same sample.
We find that when background is added, the value of \tauBz increases by
$(0.022 \pm 0.009)$~ps and the value of \Dm increases by 
$(0.020 \pm 0.005)$~ps$^{-1}$, 
where the 
error is the difference in quadrature between the statistical errors
from the fit
with and without background.
We correct our final results in data for these biases, which are each
roughly the same size as the statistical error on the results in data.
We conservatively apply a systematic uncertainty on this  bias
equal to the full statistical error on the measured result in Monte
Carlo simulation with background:  $\pm 0.018$~ps for \tauBz 
and $\pm 0.012$~ps$^{-1}$ for 
\Dm. 

\subsection{Cross-checks in data}
\label{sec:datachecks}

We perform the full maximum-likelihood fit on different subsets
of the data and find no statistically significant difference in the
results for different subsets.
The fit is performed on datasets divided according to tagging category, 
$b$-quark flavor of the \btodstlnu candidate, $b$-quark flavor of the tagging
$B$, and $D^0$ decay mode.
We also vary the range of \Dt over which we perform the fit 
(maximum value of $|\Dt|$ equal to 10, 14, and 18~ps), 
and decrease the maximum allowed value of \sigmaDt from 1.8~ps to 1.4~ps.
Again, we do not find statistically significant changes in the values of 
\tauBz or \Dm.

%%%%%%%%%%%%%%%%%%%%%%%%%%%%%%%%%%%%%%%%%%%%%

\section{Systematic Uncertainties}

\label{sec:systematics}

%% Description of systematics

We estimate systematic uncertainties on the parameters \tauBz and
\Dm with studies performed on both data and Monte Carlo samples, and
obtain the results summarized in Table~\ref{tab:systematics}. 

\begin{table}[htb]
\caption{Summary of systematic uncertainties on the two physics parameters, 
\tauBz and \Dm.}
\begin{center}
\begin{tabular*}{0.48\textwidth}{@{\extracolsep{\fill}}lcc}
\hline\hline
Source                & $\delta(\Dm)$ (ps$^{-1}$) & $\delta(\tauBz)$ (ps) \\
\hline
Selection and fit bias        & $0.0123$ & $0.0178$ \\
$z$ scale                     & $0.0020$ & $0.0060$ \\
PEP-II boost                  & $0.0005$ & $0.0015$ \\
Beam spot position            & $0.0010$ & $0.0050$ \\
SVT alignment                 & $0.0030$ & $0.0056$ \\
Background / signal prob.     & $0.0029$ & $0.0032$ \\
Background $\Delta t$ models  & $0.0012$ & $0.0063$ \\
Fixed $B^+$/$B^0$ lifetime ratio & $0.0003$ & $0.0019$ \\
Fixed $B^+$/$B^0$ mistag ratio    & $0.0001$ & $0.0003$ \\
Fixed signal outlier shape    & $0.0010$ & $0.0054$ \\
Signal resolution model       & $0.0009$ & $0.0034$ \\
\hline
Total systematic error             & $0.013$~\, & $0.022$~\,   \\
\hline\hline
\end{tabular*}
\end{center}
\label{tab:systematics}
\end{table}

The largest source of systematic uncertainty on both parameters is the
limited statistical precision for determining the bias due to the
fit procedure (in particular, the background modeling) with Monte
Carlo events. We assign the statistical errors of a full fit to Monte
Carlo samples including background to estimate this systematic uncertainty.
See Sec.~\ref{sec:MCtests} for more details.

The calculation of the decay-time difference \Dt for each event
assumes a nominal detector $z$-scale, PEP-II boost, vertical beam-spot
position, and SVT internal alignment. 
The PEP-II boost has an uncertainty of 0.1\%~\cite{ref:babar} based on
our knowledge of the beam energies. The 
$z$-scale uncertainty is determined by reconstructing protons scattered
from the beam pipe and comparing the measured beam pipe dimensions
with the optical survey data. The $z$-scale uncertainty is less than 0.4\%.
We shift the vertical beam-spot position by up to 80~$\mu$m, or vary the
position randomly with a Gaussian distribution with a width of up to
80~$\mu$m, and assign the variation in the fitted parameters as
a systematic uncertainty. The systematic uncertainty due to 
residual errors in SVT internal alignment is estimated by reprocessing the
simulated sample with different internal alignment errors. We assign
the shift of the fitted parameters as a systematic uncertainty.

The modeling of the \massdiff distribution determines
the probability we assign for each event to be due to signal.
We estimate the uncertainty due to the signal
probability calculations by repeating the full fit using an ensemble
of different signal and background parameters for the \massdiff
distributions, varied randomly 
according to the measured statistical uncertainties and  correlations 
between the parameters. We assign
the spread in each of the resulting fitted physics parameters as the
systematic uncertainty.

The modeling of the background \Dt distribution affects 
the expected background contributions to the sample.
The systematic uncertainty due to the assumed background \Dt
distributions is estimated as the shift in the fitted parameters when 
the model for the largest background (due to combinatorial-\dst events) is
replaced by the sum of a prompt term and a lifetime term.

The model of the charged $B$ background assumes fixed $B^+/\Bz$ ratios
for the mistag rates and lifetimes. 
We vary the mistag ratio by the uncertainty determined from
separate fits to hadronic $B$ decays.
We vary the lifetime ratio by the statistical uncertainty on the world 
average~\cite{ref:PDG2002}.
The resulting change in the fitted physics  parameters 
is assigned as a systematic uncertainty.

The final category of systematic uncertainties is due to assumptions
about the resolution model for signal events. We largely avoid
assumptions by floating many parameters to describe the resolution
simultaneously with the parameters of interest. However, two sources
of systematic uncertainty remain: the shape of the outlier
contribution, which cannot be determined from data alone, and the
assumed parameterization of the resolution for non-outlier
events. We study the sensitivity to the outlier shape by repeating the
full fit with outlier Gaussian functions of different
means and widths. The mean is varied between $-1$~ps and $-10$~ps, and
the width is varied from 4~ps to 12~ps.
We assign the spread
of the resulting fitted values as a systematic uncertainty. 
We estimate the uncertainty due to the assumed resolution
parameterization by repeating the full fit with a triple-Gaussian
resolution model and assigning the shift in the fitted values as the
uncertainty.

The total systematic uncertainty on \tauBz is 0.022~ps and on
\Dm is 0.013~ps$^{-1}$.

%%%%%%%%%%%%%%%%%%%%%%%%%%%%%%%%%%%%%%%%%%%%%

\section{Summary}

\label{sec:Summary}

We use a sample of approximately 14,000 exclusively reconstructed
$B^0 \rightarrow D^{*-}\ell^+\nu_\ell$ signal events to simultaneously
measure the $B^0$ lifetime $\tauBz$ and oscillation frequency $\Dm$.
We also use samples of events enhanced in the major
types of backgrounds.
The lifetime and oscillation frequency are determined 
with an unbinned maximum-likelihood fit that uses, 
for each event, the measured difference in decay times of the two $B$
mesons and its uncertainty,
the signal and background probabilities,
and $b$-quark tagging information for the other $B$.
In addition to the lifetime and oscillation frequency,
we extract the parameters describing the signal $\Dt$ resolution
function, the background $\Dt$ models, the mistag fractions,
and the $B^+$ background fraction, in
the simultaneous fit to signal and background samples. 
The results for the physics parameters are 
$$
\tauBz = (1.523^{+0.024}_{-0.023} \pm 0.022)~\rm{ps}
$$
and
$$
\Dm = (0.492 \pm 0.018 \pm 0.013)~\rm{ps}^{-1}.
$$
The statistical correlation coefficient between \tauBz\ and \Dm\ is $-0.22$.

Both the lifetime and mixing frequency have combined statistical and 
systematic uncertainties that are comparable to those of the most
precise previously-published experimental measurements~\cite{ref:PDG2002}.
The results are consistent with the world average measurements of 
$\tauBz = (1.542\pm 0.016)~\rm{ps}$
and
$\Dm = (0.489 \pm 0.008)~\rm{ps}^{-1}$~\cite{ref:PDG2002}.

This analysis demonstrates the feasibility of measuring
the $B^0$ lifetime and mixing frequency simultaneously at
$B$ Factory experiments, realizing the advantages of better
determinations of the $\Delta t$ resolution function and
the amount of $B^+$ background.
All background fractions, $\Delta t$ resolution parameters 
for signal and background, and mistag fractions 
are determined from the data.
The lifetime is most correlated with the $\Delta t$ resolution
parameters for signal, while the mixing frequency is most
correlated with the $B^+$ background fraction.
The largest systematic uncertainty on both parameters is
the limited statistical precision for determining any bias
due to the fit procedure (in particular, the background modeling)
with Monte Carlo simulation.

Both the statistical and systematic uncertainties on these
parameters can be reduced with the larger data and Monte
Carlo simulation samples already available at the $B$ Factories.
Other physics parameters, such as the difference in decay rates
of the neutral $B$ mass eigenstates, can also be included
in a simultaneous fit in future data samples.

%%%%%%%%%%%%%%%%%%%%%%%%%%%%%%%%%%%%%%%%%%%%%

\section{Acknowledgments}

\label{sec:Acknowledgments}

We are grateful for the 
extraordinary contributions of our \pep2\ colleagues in
achieving the excellent luminosity and machine conditions
that have made this work possible.
The success of this project also relies critically on the 
expertise and dedication of the computing organizations that 
support \babar.
The collaborating institutions wish to thank 
SLAC for its support and the kind hospitality extended to them. 
This work is supported by the
US Department of Energy
and National Science Foundation, the
Natural Sciences and Engineering Research Council (Canada),
Institute of High Energy Physics (China), the
Commissariat \`a l'Energie Atomique and
Institut National de Physique Nucl\'eaire et de Physique des Particules
(France), the
Bundesministerium f\"ur Bildung und Forschung and
Deutsche Forschungsgemeinschaft
(Germany), the
Istituto Nazionale di Fisica Nucleare (Italy),
the Research Council of Norway, the
Ministry of Science and Technology of the Russian Federation, and the
Particle Physics and Astronomy Research Council (United Kingdom). 
Individuals have received support from 
the A. P. Sloan Foundation, 
the Research Corporation,
and the Alexander von Humboldt Foundation.

%%%%%%%%%%%%%%%%%%%%%%%%%%%%%%%%%%%%%%%%%%%%%


\begin{thebibliography}{99}
\bibitem{cc}
Throughout this 
paper, charge conjugate modes are always implied.
\bibitem{vtxBs}
This improvement
is estimated with a large Monte Carlo simulation of signal events.
\end{thebibliography}
\end{document}